\documentclass[journal]{IEEEtran}
\usepackage{bm}
\usepackage{physics}
\usepackage{hyperref}
\usepackage{cite}
\usepackage{amsmath,amssymb,amsfonts}
\usepackage{graphicx}
\usepackage{textcomp}
\usepackage{wrapfig}
\usepackage[latin9]{inputenc}
\usepackage{float}
\usepackage{algorithm}   
\usepackage{algpseudocode}
\usepackage{pbalance}
\usepackage{cite}
\usepackage{xcolor}
\usepackage{graphicx}
\usepackage{booktabs}
\usepackage{multirow}
\usepackage{array}
\usepackage{siunitx}
\algtext*{EndFor}
\algtext*{EndIf}
\algtext*{EndProcedure}
\usepackage{xurl} 

\makeatletter
\g@addto@macro\normalsize{%
	\setlength{\abovedisplayskip}{6pt plus 3pt minus 2pt}%
	\setlength{\belowdisplayskip}{6pt plus 3pt minus 2pt}}
\makeatother

\begin{document}

\title{Unrolled RF Holographic Imaging: Structured Sparsity and Low-Rank EM Model Adaptation}
\author{Federica Fieramosca, \IEEEmembership{Member, IEEE}, Alexander H. Paulus, \IEEEmembership{Member, IEEE}, Richard Oliveira, \IEEEmembership{Student Member, IEEE}, Stefano Savazzi, \IEEEmembership{Senior Member, IEEE} \thanks{\protect \\Funded by the European Union (EU) Pathfinder Open project HOLDEN (Ethical Design of Holography in Dense Wireless Networks) under Grant Agreement (GA) no. 101099491,  and by European Marie Sk\l{}odowska-Curie Actions (MSCA) Doctoral Network SMARTTEST under GA no. 101167834.\protect \\Federica Fieramosca and Alexander Paulus are co-first authors. \\F. Fieramosca R. Oliveira and S. Savazzi are with the Consiglio Nazionale delle Ricerche, IEIIT institute, 20133 Milan, Italy (e-mail: name.surname@cnr.it). \\A. Paulus is with the Technical University of Munich (TUM), Germany (e-mail: a.paulus@tum.de).}}

\maketitle
	
\begin{abstract}
Radio-Frequency (RF) holographic imaging reconstructs a volumetric map of the permittivity contrast from phase-coherent samples of the scattered electromagnetic (EM) field. The resulting inverse problem is severely ill-posed, as the receivers are orders of magnitude fewer than the unknown 3D volume elements (voxels). It is classically regularized by sparsity-promoting solvers such as the Iterative Shrinkage-Thresholding Algorithm (ISTA). Two assumptions limit these solvers: the $\ell_1$ penalty is spatially uniform, and the EM forward model is typically approximated as a linear operator. This paper revisits the problem through \emph{algorithm unrolling}, in which the solver iterations become the layers of a compact trainable network that preserves the EM forward model and learns only a few interpretable parameters from limited data. Building on the Learned ISTA (LISTA), a Weighted LISTA (W-LISTA) is first proposed, which learns a spatially-varying $\ell_1$ regularization, steering the sparsity prior towards target shapes consistent with the deployment. Second, the Low-Rank Weighted LISTA (LoRaW-LISTA) applies a low-rank adaptation (LoRa) of the holographic operator to compensate for model mismatches from linearized EM approximations. Both methods are validated on full-wave EM simulations and on a $2.45$\,GHz indoor measurement campaign with human-body phantoms. Combining the spatially-varying regularization with the low-rank adaptation of the EM model improves the signal-to-clutter ratio by about \textcolor{black}{$70\%$ over the ISTA and LISTA baselines}, recovering structural details where classical iterative solvers fail. Inference takes less than $30$\,s to reconstruct $1$\,m$^3$ of scene on conventional GPUs. 
The proposed tools are rapidly adaptable building blocks for the sensing layer of emerging smart radio environments.
\end{abstract}

\section{Introduction}

Radio-frequency (RF) holography is an emerging sensing modality that
reconstructs a three-dimensional (3D) representation of a monitored
region from phase-sensitive samples of the electromagnetic (EM) field collected over a large scale receive antenna array. Early Wi-Fi demonstrations showed that phase-coherent RF measurements can
recover 3D information about the surrounding
environment~\cite{holl2017holography,amineh2011three}, building on a
long-standing line of research on device-free wireless sensing and radio
tomographic imaging~\cite{tomog,FieramoscaIoTJ2025}.

\begin{figure}[t]
	\centering
	\includegraphics[width=\linewidth]{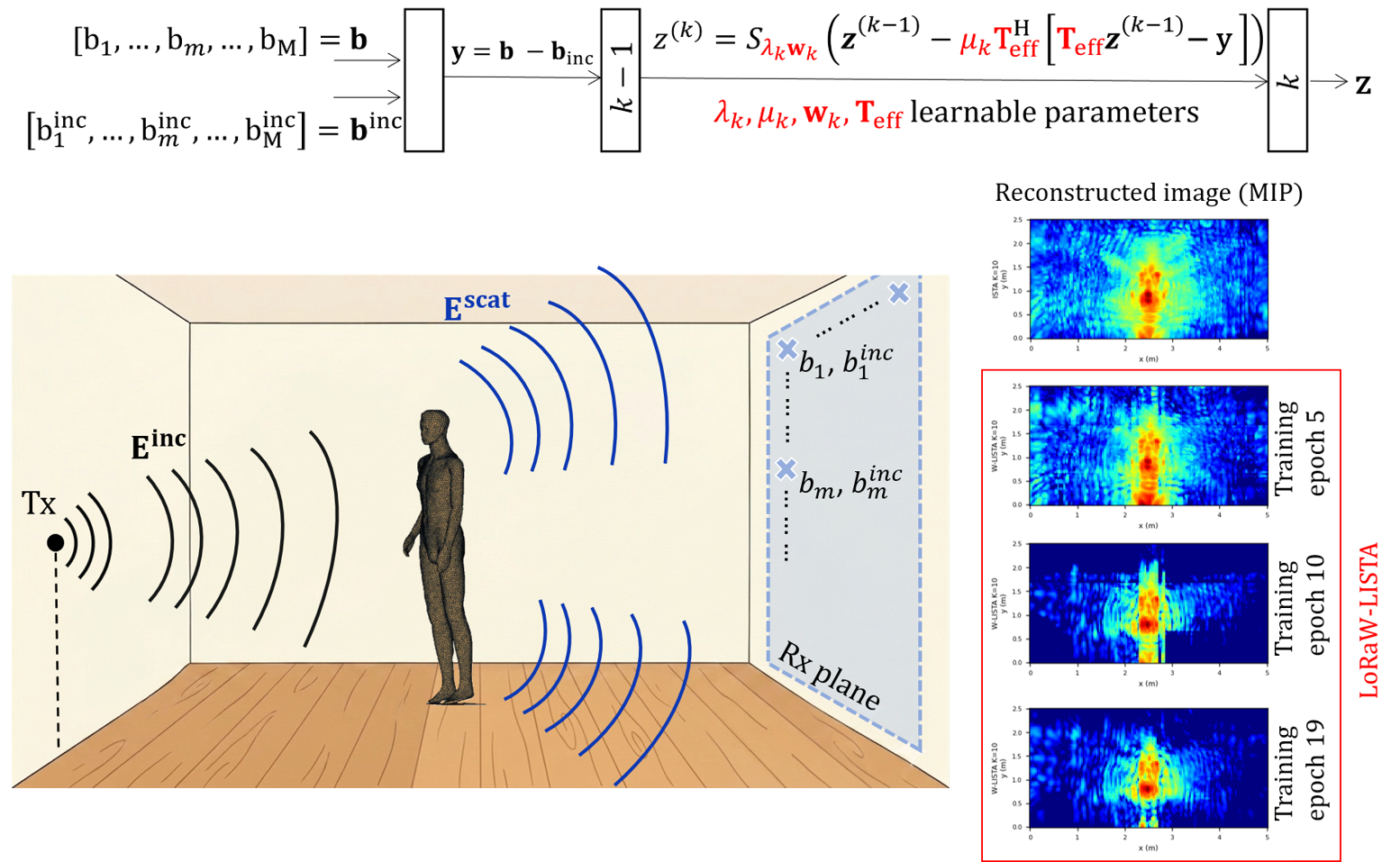}
	\caption{Overview of the RF holographic imaging pipeline. The scattered field $\mathbf{y}=\mathbf{b}-\mathbf{b}^{\mathrm{inc}}$, sampled over the RX plane, is inverted by an unrolled network that keeps the EM operator fixed and learns only a few interpretable parameters ($\mu_k,\lambda_k,\mathbf{w}_k, \mathbf{T}_{\mathrm{eff}}$). The reconstruction sharpens on the body support across consecutive training epochs (right).}
	\label{fig:scenario_meas_scheme}
\end{figure}

Holographic processing has long been employed at microwave and millimeter-wave frequencies, where it underpins near-field imaging of concealed objects in security screening~\cite{Sheen2001,Ahmed2011},
non-destructive testing and biomedical imaging~\cite{Nikolova2017}, and synthetic-aperture and ground-penetrating radar~\cite{Soldovieri2009}.
More recently, the densification of antenna arrays, reconfigurable intelligent
surfaces and cooperative wireless infrastructures, often referred to as Smart
Radio Environments (SREs)~\cite{savazzi2026holography}, have enabled
holographic imaging at a few GHz, where penetration through walls and
large-scale coverage are achievable, at the price of coarser angular and depth
resolution~\cite{adib2013see,adib2015capturing,wang2019survey}.
Massive-MIMO arrays improve the localization accuracy of passive targets
through their larger aperture~\cite{zeng2021massivemimo}, in line with integrated sensing and communication (ISAC)
paradigms~\cite{adhikary2024holographic}. At system level, recent Wi-Fi deployments have demonstrated through-wall detection
and coarse imaging of people in the sub-6~GHz range~\cite{Sun.2021,Na.2025}, entering into
standardization process with the IEEE 802.11bf~\cite{du2025overview,wu2022ieee80211bf}.

Fig.~\ref{fig:scenario_meas_scheme} depicts the holographic imaging pipeline considered in this paper. A transmitter illuminates the monitored volume, and a large-scale planar array of receivers samples the resulting EM field. The scattered field is then inverted to recover a 3D map of the scene. Unlike conventional image-based sensing, RF holography does not produce a recognisable visual scene, but a coarse map of the scattering induced by objects and human bodies~\cite{savazzi2026holography}. Such a map
conveys an infrastructural snapshot of the environment \cite{moral}: namely, the position and shape of the targets, which can be used for localization and posture analysis. Reconstruction  quality depends on the accuracy of the EM model and on the prior knowledge used to regularize the solution. 

\subsection{Model-based holographic reconstruction with RF signals}

A standard RF holography pipeline is built around three building
blocks. The first is the EM forward model, which maps the unknown permittivity contrast inside the monitored volume into the sampled scattered field. In most practical deployments, the \emph{holographic operator} mapping is a
matrix obtained by linearizing the volume
integral equation under the first-order Born approximation, with the dyadic Green's function~\cite{Tai.1994} as
propagation kernel~\cite{Paulus.2018b,paulus2025accuracy,Pastorino.2010}. 
The second block is the inverse problem solver. Since the receivers are orders
of magnitude fewer than the 3D volume elements (voxels), the inversion is strongly ill-posed and
must be regularized. Classical solutions range from matched-filter (MF)
back-projection and Tikhonov regularization to compressive-sensing
formulations~\cite{Donoho2006,Candes2006}, typically solved by Iterative Shrinkage-Thresholding
Algorithms~\cite{Daubechies2004} (ISTA, FISTA~\cite{BeckTeboulle2009}) or by
greedy methods~\cite{Tropp2007}. The third block projects the volumetric
reconstruction onto an interpretable 2D image, via a Maximum
Intensity Projection (MIP).

Each block relies on hand-tuned parameters, i.e., the spatial
discretization, the regularization weight, the solver step size and the
post-processing threshold. Being chosen for analytical tractability and
worst-case convergence rather than for a specific deployment, they rarely
transfer across changes in room geometry, sensor layout or targets. 
Data-driven methods can partly bridge this gap~\cite{paulus2025accuracy}, but
most existing approaches replace the analytical forward model with a generic
black-box network, sacrificing physics consistency and interpretability.

These limitations motivate a holographic reconstruction pipeline that can be
adapted on a per-deployment basis from limited calibration data or coarse
prior information on the environment (i.e., a floor plan or a CAD model),
through a compact set of physically meaningful parameters.
\textcolor{black}{Such adaptability also yields direct control over the
resolution and the level of detail of the
reconstruction~\cite{savazzi2026holography}.}

\subsection{Algorithm unrolling: state of the art and open issues}

\textcolor{black}{Algorithm unrolling provides a principled bridge
between model-based optimization and data-driven inference~\cite{spm}: a fixed number of solver iterations is mapped onto the layers of a
feed-forward network, and the step sizes, thresholds and regularization
parameters are learned end-to-end from a labelled dataset. Since the introduction of Learned ISTA (LISTA)~\cite{Gregor2010,Chen2018} and its approximate
message passing counterpart~\cite{Borgerding2017}, the paradigm has become an established instance of model-based deep
learning~\cite{Shlezinger2023}, applied in adaptive
filtering~\cite{Hou2026SPL}, graph learning~\cite{Zhang2025TSP}, array
processing for ISAC~\cite{Wang2026ICASSP} and tomographic image
reconstruction~\cite{Zhang2025EIT}.}

\textcolor{black}{The applicability of unrolling tools to the holographic inverse problem is limited by three main challenges. First, most unrolled
architectures learn the measurement operator itself, replacing it with
unconstrained weight matrices~\cite{Gregor2010,Borgerding2017}. Such a choice is incompatible with operators derived from physics (i.e.,
RF propagation), which may be too large to be stored explicitly and are
typically applied on-the-fly. Physics-informed reconstructions are thus required to maintain consistency with the underlying EM model. Second, the sparsity priors currently proposed are spatially homogeneous, being parametrized by a single threshold per unrolled layer. Richer regularizers have been unrolled recently, either learned from data~\cite{Chen2025TII, unser2026universal} or built on structured signal models such as graph Laplacians~\cite{Cai2026TIP}, but they characterize the signal class rather than the geometry of the reconstructed scene. In holographic imaging, the scatterers occupy a small and spatially coherent
portion of the monitored volume.
The sparsity prior should therefore be
spatially varying and described by a small number of
parameters. Third, the forward operator is assumed to be exact, whereas in practice it is an approximate description of the true scattering
process: the resulting bias
affects every layer of the unrolled network. A structured (spatial)
prior that resolves the spatial support of the scene, and a low-rank correction of the model mismatch constitute the main focus of this work.}

\subsection{Contributions}

The paper introduces novel algorithm unrolling~\cite{spm} formulations
tailored for the RF holographic inverse problem. The model-based
iterative solver is reinterpreted as a differentiable deep network
whose hyper-parameters are learned end-to-end. \textcolor{black}{Two novel unrolled
algorithms are proposed, addressing respectively the spatial structure of the
unknown support and the mismatch of the linearized forward model, and are of
interest beyond the holographic setting.} To the best of the authors'
knowledge, this is the first application of algorithm unrolling to large-scale
3D indoor RF holographic reconstruction. The contributions of this work are
fourfold.
\begin{itemize}[]
\item A novel Weighted LISTA (W-LISTA) architecture, extending standard
Learned ISTA~\cite{Gregor2010,Chen2018} (LISTA), is introduced to regularize
the reconstruction of the 3D scene. W-LISTA replaces the $\ell_1$ penalty of
ISTA/LISTA with a learnable, spatially-varying $\ell_1$ regularization,
namely a per-voxel \emph{spatial prior} (sparsity prior) that steers the reconstruction towards the
3D regions and target shapes consistent with the deployment. \textcolor{black}{A
rank-1 separable factorization along the Cartesian axes keeps the number of
learnable parameters compact even for large volumes}, while the site-specific
prior knowledge of the environment is embedded directly into the imaging
reconstruction process.

\item A physics-informed low-rank correction of the holographic operator is
further proposed as inspired by low-rank adaptation (LoRA) of large
pre-trained models~\cite{hu2021lora}. The proposed Low-Rank Weighted LISTA
(LoRaW-LISTA) extends the W-LISTA compensating for the model mismatch
introduced by the linearized EM approximations for certain target types.
\textcolor{black}{The holographic operator is never learned as only a low-rank correction is estimated, jointly with the per-layer step sizes, base thresholds and spatial weights.}

\item A supervised procedure is developed to adapt the unrolled algorithms
hyper-parameters on a per-deployment basis from EM-simulated or measured
calibration data, optionally complemented by prior information on the
environment (i.e., a floor plan map or CAD model of the building/room).
\textcolor{black}{A two-stage schedule is adopted, in which the sparsity prior is
learned first and the low-rank operator correction is enabled afterwards:
correction thus targets the residual model mismatch rather than being affected by the
regularization error.}

\item The framework is validated in two settings: (i)~full-wave EM simulated
environments, which yield fully labelled training data with arbitrary scenes
and geometries, and (ii)~an indoor case study with laboratory measurements at
$2.45$\,GHz and human-body phantoms, probing robustness to calibration error
and unmodelled multipath components. LoRaW-LISTA and W-LISTA are benchmarked
against state-of-the-art MF back-projection and $\ell_1$
regularization (ISTA), and plain LISTA.
\end{itemize}

The paper is organized as follows:
Sections~\ref{sec:rf_holography}--\ref{sec:unrolling} review the EM forward
model, the classical reconstruction pipeline and the LISTA baseline,
Sections~\ref{sec:wlista}--\ref{sec:lrwlista} present the two proposed
architectures, and Sections~\ref{sec:simulations}--\ref{sec:indoor} report the
validation on EM simulations and the indoor case study.

\section{RF Holography Problem}
\label{sec:rf_holography}
	
The section discusses the signal processing framework for RF holographic imaging based on EM forward modeling and inverse scattering reconstruction. The theoretical basis combines classical wave propagation theory, linear inverse problems, and sparse regularization methods.
	
We refer to holographic imaging as the recovery of an unknown material-contrast distribution within a 3D domain $\Omega$ from scattered electromagnetic fields sampled on a 2D observation surface with RF transceivers, as illustrated in Fig.~\ref{fig:scenario_meas_scheme}. 
We define the total field at an observation point $\bm r$ as the superposition of an incident field and a scattered field, $\mathbf E(\mathbf r)=\mathbf E^{\mathrm{inc}}(\mathbf r)+\mathbf E^{\mathrm{scat}}(\mathbf r)$, where the incident field is that of the homogeneous medium with background permittivity $\varepsilon_b$. The scattered field is the perturbation induced by objects with relative permittivity contrast $\chi(\mathbf r)=\varepsilon_r(\mathbf r)-\varepsilon_b$; in the following, we assume the background is vacuum, $\varepsilon_b=1$.
	
The propagation of RF electromagnetic fields in a medium with spatially varying permittivity $\varepsilon(\mathbf r) = \varepsilon_r(\bm r)\varepsilon_0$ is governed by
		\begin{equation}
		\nabla^{2} \mathbf E(\mathbf r) 
		- k_0^2 \varepsilon_r(\mathbf r)\mathbf E(\mathbf r) = 0,\label{eq:inhomogeneous_helmholtz}
	\end{equation}
where $k_0 = \frac{\omega}{c}$, $\omega = 2\pi f_0$, $f_0$ being the operating frequency, $\epsilon_0$ being the vacuum permittivity and $\varepsilon_r(\mathbf r)$ as the relative permittivity distribution. 
A solution to~\eqref{eq:inhomogeneous_helmholtz} is nontrivial and would require the knowledge of the appropriate Green's function of the inhomogeneous material distribution. However, using the volume equivalence principle~\cite{Jin.2015}, the inhomogeneous medium can be accounted for by an equivalent volumetric current density	
	\begin{equation}
		\mathbf J(\mathbf r)
		=
		\mathrm{j}\omega\epsilon_0
		\chi(\mathbf r)
		\mathbf E(\mathbf r).
	\end{equation}
The electric field can therefore be expressed as
	\begin{equation}
		\mathbf E(\mathbf r)
		=
		\mathbf E^{\mathrm{inc}}(\mathbf r)
		+
		k_0^2\int_{\Omega}
		\overline{\mathbf G}(\mathbf r,\mathbf r')\cdot
		\chi(\mathbf r')
		\mathbf E(\mathbf r')
		d\mathbf r'
		\label{forw}
	\end{equation}
    using the dyadic Green's function~\cite{Paulus.2018b,Tai.1994}
	\begin{equation}
		\overline{\mathbf G}(\mathbf r,\mathbf r')
		=
		\left(
		\overline{\mathbf I}
		+
		\frac{1}{k^2}\nabla^{2}
		\right)
		\frac{\mathrm{e}^{-\mathrm{j}k|\mathbf r-\mathbf r'|}}
		{4\pi |\mathbf r-\mathbf r'|}
	\end{equation}
	of free space, where $\mathbf r$ and $\mathbf r'$ denote the observation (field) point and the source point, respectively.

\subsection{Linearized scattering model and 3D voxel discretizations}
The continuous formulation \eqref{forw} is now specialized to a discrete geometry assuming a vertical polarization. The imaging domain is discretized into $N$ 3D voxels $V_n$ according to
	\begin{equation}
		\Omega
		=
		\bigcup_{n=1}^{N}
		V_n
	\end{equation}
with barycenter locations $\mathbf{r}_n$ and volume $\Delta V_n$. The contrast distribution is now represented as the discretized vector
	\begin{equation}
		\mathbf z
		=
		[\chi(\mathbf{r}_1),\dots,\chi(\mathbf{r}_N)]^{\mathrm{T}}.
		\label{contrast}
	\end{equation}
The scattered field is sampled at $M$ elements of a large-scale planar receive array arranged on a regular 2D grid, with antenna $m$ located at $\mathbf r_m$.

A linearized model follows from the first-order Born approximation \cite{Born.1999}, setting $\mathbf E(\mathbf r_n)\approx\mathbf E^{\mathrm{inc}}(\mathbf r_n)$ inside $\Omega$. Evaluating \eqref{forw} at the antenna positions $\mathbf r_m$ and discretizing the volume integral over the $N$ voxels yields the linear forward model
	\begin{equation}
		\mathbf E^{\mathrm{scat}}(\mathbf r_m)
		=k_0^2
		\sum_{n=1}^{N}
		\overline{\mathbf G}(\mathbf r_m,\mathbf r_n)
		\,\chi(\mathbf r_n)\,
		\mathbf E^{\mathrm{inc}}(\mathbf r_n)
		\,\Delta V_n . \label{born}
	\end{equation}
The approximation is known to be valid when the scattering objects exhibit moderate permittivity contrast and when multiple scattering effects are limited.

At antenna $m$, the measured transmission coefficient $S_{21}(\mathbf r_m)$ provides a sample of the total field, while $S_{21}^{\mathrm{inc}}(\mathbf r_m)$ is the corresponding measurement of the incident (background) field acquired without targets. Stacking these samples into the vectors $\mathbf b=[b_m]_{m=1}^M$ and $\mathbf b^{\mathrm{inc}}=[b_m^{\mathrm{inc}}]_{m=1}^M$, with $b_m\propto S_{21}(\mathbf r_m)$ and $b_m^{\mathrm{inc}}\propto S_{21}^{\mathrm{inc}}(\mathbf r_m)$, the sampled scattered field is obtained as the calibrated difference
	\begin{equation}
		\mathbf y
		= [\mathbf E^{\mathrm{scat}}(\mathbf r_1),...,\mathbf E^{\mathrm{scat}}(\mathbf r_M)] \approx
		\mathbf b - \mathbf b^{\mathrm{inc}}
		\in
		\mathbb C^M.
		\label{eq:scat_meas}
	\end{equation}
Finally, using \eqref{contrast} and \eqref{born}, for $m=1,\dots,M$ it yields the compact discrete linear forward model
	\begin{equation}
		\mathbf y
		=
		\mathbf T \mathbf z.
		\label{eq:linear_model}
	\end{equation}
The entries of the holographic operator $\mathbf T$ are
	\begin{equation}
		T_{mn}
		=
		k_0^2\overline{\mathbf G}(\mathbf r_m,\mathbf r_n)
		\mathbf E^{\mathrm{inc}}(\mathbf r_n)\Delta V_n
		\label{tomographic_operator}
	\end{equation}
and map the contrast of voxel $n$ into the field measured at antenna $m$, namely the wave propagation between voxels and receiving antennas. The incident field is assumed to be a uniform plane wave over the imaging domain, i.e., $\mathbf{E}^{\mathrm{inc}}(\mathbf{r}_n)=1$ for all voxels $n$.

	
	\subsection{Regularization and Sparse Reconstruction}
	
	The inverse problem is strongly ill-posed since $M \ll N$ and regularization is therefore required. The spatial sampling density of the array elements on the 2D wall should satisfy $\Delta x \lesssim \frac{\lambda}{2}$ to avoid spatial aliasing. Notice that from a signal processing perspective, the forward operator $\mathbf T$ acts as a propagation kernel,  the reconstruction corresponds to a deconvolution problem over a spatial sampling grid, the sparsity regularization introduces robustness to noise, and the Green function acts as a spatial impulse response.

	A simplified solution to~\eqref{eq:linear_model} can be obtained using MF, or back-projection, via
		\begin{equation}
		\hat{\mathbf z}_{\mathrm{MF}}
		=
		\mathbf T^{\mathrm{H}} \mathbf y,
        \label{eq:mf}
	\end{equation}
where $(\cdot)^{\mathrm{H}}$ denotes the Hermitian transpose.
	This operation corresponds to correlating the measured field with the Green function associated with each voxel location. The reconstruction provides a fast approximation of the inverse solution, whose resolution is determined by the spatial bandwidth of the measurement configuration. 
    Practical implementations often include the normalization ($T_{mn}^{*}$ refers to complex conjugation)	
	\begin{equation}
		\hat z_n
		=
		\frac{
			\sum_m
			T_{mn}^{*} y_m
		}{
			\sum_m
			|T_{mn}|^2
		},\label{eq:normalization}
	\end{equation}
	to compensate for spatially varying illumination and improve image contrast.

	In this paper we target a sparsity-promoting reconstruction defined as
	\begin{equation}
		\hat{\mathbf z}
		=
		\arg\min_{\mathbf z}
		\left\|
		\mathbf T \mathbf z - \mathbf y
		\right\|_2^2
		+
		\alpha
		\|\mathbf z\|_1
		\label{l1_reg},
	\end{equation}
	where $\alpha$ controls the sparsity level. This formulation assumes that scattering is spatially sparse, i.e., only few targets are co-present. A solution to \eqref{l1_reg} can be obtained by ISTA.
	\subsection{Holographic imaging via Maximum Intensity Projection}
	
	The inversion procedure produces a volumetric estimate of the scattering intensity $\hat {\mathbf{z}}(x,y,z)$	defined over the 3D imaging domain, where $x$ denotes the horizontal coordinate along the receive array plane (surface), $y$ denotes the vertical coordinate (height, positive downward), $z$ represents the depth inside the monitored environment.
	
	To obtain a 2D representation of the scene, the reconstructed 3D intensity is projected along the depth direction using a Maximum Intensity Projection (MIP) defined as	
	\begin{equation}
		I(x,y)
		=
		\max_{z}
		\left|
		\hat {\mathbf{z}}(x,y,z)
		\right|
		\label{mip}.
	\end{equation}
This operation produces a 2D image defined on the receiver plane which highlights the most significant scattering contributions observed as orthogonal to the measurement surface. The resulting map lies on the lateral-height $(x,y)$ plane, with the horizontal axis Tx-centered and the vertical axis reporting height, consistently with the reference frame of Fig.~\ref{fig:meas_scenario}.
	
The MIP operation can be interpreted as selecting, for each spatial coordinate $(x,y)$, the voxel that exhibits the strongest scattering response along the depth direction. Utilizing the normalization in~\eqref{eq:normalization}, the resulting 2D map represents the projected reflectivity of the scene as $I(x,y)\approx\max_{z}\left|\chi(x,y,z)\right|$.

\section{Reconstruction via Unrolled Deep Networks}
\label{sec:unrolling}

Algorithm unrolling~\cite{Gregor2010,Chen2018} offers a principled way
to keep the holographic EM forward model at the core of the inversion problem as in \eqref{l1_reg} while making every other element of the pipeline trainable.
The core idea is to interpret a fixed number of iterations of a model-based iterative solver as the layers of a recursive deep network (or a computational graph), expose its internal hyper-parameters, namely the
step sizes, thresholds, per-voxel priors, frequency weights, as
free parameters, and learn them by backpropagation. Supervised training of such hyper-parameters can be based on
(i)~EM-simulated training samples obtained from the same Green's
function used at inference time, or 
(ii)~calibration measurements acquired on the actual deployment site,
possibly together with side information on the relative permittivity
contrast of the environment (i.e., from a known floor-plan, a CAD
model, or a coarse occupancy prior).

The proposed unrolling approach learns how to build a regularization environment that uses the propagation operator $\mathbf{T}$ from~\eqref{tomographic_operator} more efficiently, yielding a pipeline that is physics-consistent by construction, namely every layer
respects the spatial impulse response of the Green function and is interpretable, since the learned parameters are exactly the per-iteration step sizes and
shrinkage thresholds of the underlying solver. It is also scalable, requiring only $K\!\sim\!10$ layers against the hundreds of iterations needed by plain ISTA, and flexible, as the same architecture is retrained
for new sites or sensor layouts without changing the analytical EM model.

The unrolled network can be supervised either with EM-simulated samples or with prior information on the environment permittivity contrast. Section~\ref{sec:ista} revisits ISTA, the iterative solver
of~\eqref{l1_reg} adopted as backbone of the unrolled network.
Section~\ref{sec:lista} then describes LISTA,
the canonical unrolling of ISTA in which the per-iteration step size
and shrinkage threshold are turned into trainable parameters.

\subsection{Iterative Shrinkage-Thresholding Algorithm (ISTA)}
\label{sec:ista}

Problem in~\eqref{l1_reg} is commonly solved via ISTA. Specifically, 
given an initialization $\mathbf{z}^{(0)}$, each iteration performs
a gradient descent step on the data-fidelity term followed by a proximal
operator which enforces sparsity. The solution at the $(k+1)$-th iteration is then given by
\begin{equation}
	\mathbf{z}^{(k+1)}
	=
	\mathcal{S}_{\lambda}
	\!\left(
	\mathbf{z}^{(k)}
	-
	\tau \,\mathbf{T}^{\mathrm{H}}
	\!\left(\mathbf{T}\mathbf{z}^{(k)} - \mathbf{y}\right)
	\right),
	\label{eq:ista}
\end{equation}
where $\tau = s/L$ is the step size, $s \leq 1$ a scaling
factor, and $L$ is the Lipschitz constant of $\mathbf{T}^{\mathrm{H}}\mathbf{T}$ which is estimated
via power iteration. Moreover, $\mathcal{S}_\lambda(\cdot)$ is the \emph{proximal operator}, known as the \emph{soft-thresholding} operator, which is given by
\begin{equation}
	\mathcal{S}_\lambda(u)
	=
	\frac{u}{|u|}
	\max\!\left(|u| - \lambda,\, 0\right),
\end{equation}
where $\lambda = \tau \alpha$ is the  shrinkage threshold. The parameters $\tau$ and $\alpha$ are fixed scalars applied identically at every iteration. Notice that the physical operator $\mathbf{T}$ is never stored explicitly (as unfeasible given its size) but remains fixed and is evaluated
on-the-fly using~\eqref{tomographic_operator}.

Convergence of ISTA is guaranteed when $\tau \leq 1/L$, but the
worst-case step size bound is often conservative, requiring several iterations $N_\mathrm{iter}$ to reach an acceptable solution.

\subsection{Algorithm unrolling and Learned ISTA (LISTA)}
\label{sec:lista}

Learned ISTA (LISTA)~\cite{Gregor2010}, ~\cite{cherkaoui2020learning} transforms the iterations~\eqref{eq:ista} into a \emph{fixed computational
	graph} of depth $N_\mathrm{iter}$, whose scalar hyper-parameters
$(\tau, \alpha)$ can be tuned by hand. Let $\{(\mathbf{y}_i,\, \mathbf{z}_i^*)\}_{i=1}^{P}$ be a dataset of $P$ labelled pairs, where $\mathbf{z}_i^*$ denotes the information about true permittivity contrast in some regions of interest (ROI) $i$ and $\mathbf{y}_i$ is a corresponding set of EM field measurements collected either from a synthetic environment \cite{eucap26} or a real setup, the step sizes and thresholds $\mu_k$ and $\lambda_k$ can be \emph{learned}
by treating each iteration as a layer of a recursive neural network.

Specifically, LISTA \emph{unrolls} $K$ iterations of~\eqref{eq:ista},
assigning an independent learnable pair $(\mu_k, \lambda_k)$ to each
neural network layer $k = 1, \dots, K$ such that
\begin{equation}
	\mathbf{z}^{(k)}
	=
	\mathcal{S}_{\lambda_k}
	\!\left(
	\mathbf{z}^{(k-1)}
	-
	\mu_k\,
	\mathbf{T}^{\mathrm{H}}
	\!\left(\mathbf{T}\mathbf{z}^{(k-1)} - \mathbf{y}\right)
	\right),
	\label{eq:lista}
\end{equation}
where the projection $\mathcal{S}_{\lambda_{k}}(\cdot)$ corresponds to a neural network non-linearity (activation function). Moreover, we set \mbox{$\mathbf{z}^{(0)} = \mathbf{T}^{\mathrm{H}} \mathbf{y}$}, namely the MF solution \eqref{eq:mf}, as a warm start.


Given a training set, the network parameters
\mbox{$\bm{\theta} = \{\mu_k, \lambda_k\}_{k=1}^K$}
are learned via 
\begin{equation}
	\hat{\bm{\theta}}
	=\arg\min_{\bm{\theta}}\mathcal{L}(\bm{\theta})=
	\arg\min_{\bm{\theta}}
	\frac{1}{P}
	\sum_{i=1}^{P}
	\bigl\|
	|\hat{\mathbf{z}}_i(\bm{\theta})| - |\mathbf{z}_i^*|
	\bigr\|_2^2,
	\label{eq:lista_loss}
\end{equation}
i.e., by minimizing the reconstruction loss $\mathcal{L}(\bm{\theta})$, namely the magnitude Mean Squared Error (MSE). Here, $\hat{\mathbf{z}}_i(\bm{\theta})$ denotes the output of the
$K-$layer neural network~\eqref{eq:lista} for input $\mathbf{y}_i$, while the
loss can be computed on the modulus, to be insensitive to global phase
ambiguity. Gradients with respect to $\bm{\theta}$ are computed by
backpropagation through the layers of the network, namely the unrolled iterations, exploiting the
linearity of $\mathbf{T}$ and $\mathbf{T}^{\mathrm{H}}$ operators.

Unlike standard ISTA, which uses a single conservative step size derived
from the worst-case Lipschitz bound, LISTA learns a problem-specific iteration step-size schedule\footnote{A typical learning rate (LR) schedule can take larger steps early, namely when the iterate is far from a stable solution, and smaller, more precise steps near convergence.} $\{\mu_k\}$ and a shrinkage threshold schedule $\{\lambda_k\}$ which adapts
the sparsity level at each stage~\cite{Chen2018}. 

The total computational cost of LISTA inference is identical to
$K$ iterations of ISTA: each layer requires exactly one evaluation of
$\mathbf{T}$ and one of $\mathbf{T}^{\mathrm{H}}$, both performed in batched form. The only overhead relative to
classical ISTA is the offline training phase, which is implemented across all the labelled dataset.
The learned parameters $\bm{\theta}$ consist of $2K$ real scalars and can be directly
substituted into the ISTA loop of the existing reconstruction pipeline
without any architectural change.

\section{W-LISTA: Weighted LISTA with Factorized Spatial Priors}
\label{sec:wlista}

In this section, we describe the proposed W-LISTA, an unrolled network specifically designed to exploit the spatial structure of indoor RF holographic scenes. Standard LISTA assigns a single scalar threshold $\lambda_k$ uniformly to
\emph{all} $N$ voxels at layer $k$. Mapping to the holographic problem of Section~\ref{sec:rf_holography}, this design implicitly assumes
that the prior probability of a non-zero permittivity contrast $\chi(\mathbf r)$ is identical at every
location in $\Omega$, regardless of the known geometry of the
imaging setup or the statistics of the training scenes.
In practice, targets might occupy a restricted sub-volume of the 3D space $\Omega$
(i.e., the human-body ROIs in a considered environment),
while the vast majority of voxels are empty background.
A spatially uniform threshold ignores this structure, as it applies
the same shrinkage to target and background voxels, and the resulting
$2\!\times\! K$ learnable scalars provide too few degrees of freedom to encode it.
The proposed W-LISTA replaces the isotropic thresholding with a
$\emph{per-voxel}$ shrinkage map, factorized as a  rank-1 separable outer product along the 3D space, so that the parameter count remains
in the order of a few thousand\footnote{Parameter count is orders of magnitude smaller than
the $K\!\times\!N$ count of a fully voxel-wise prior, which would be practically unfeasible considering typical RF holographic problems.}. In addition, W-LISTA allows the deep network to encode site-specific or prior knowledge of where, inside the 3D imaging volume, body occupancies are most likely to occur.

\subsection{Per-voxel weighted $\ell_1$ proximal operator}

W-LISTA replaces the uniform $\ell_1$ penalty with a
\emph{voxel-weighted} counterpart.
At each layer $k$, the proximal step in \eqref{eq:ista} minimizes
\begin{equation}
    \mathcal{P}_k(\mathbf{z})
    =
    \left\|
    \mathrm{diag}(\mathbf{w}_k)\,\mathbf{z}
    \right\|_1
    =
    \sum_{n=1}^{N} w_{k,n}\,|z_n|,
    \label{eq:wl1}
\end{equation}
where $\mathbf{w}_k = [w_{k,1},\dots,w_{k,N}]^{\mathrm{T}} \in \mathbb{R}_{+}^N$
is a layer-specific positive weight vector.
A large weight $w_{k,n}$ imposes a strong sparsity prior on voxel $n$
(background suppression), while a small weight
relaxes shrinkage where targets are expected to be located.

The closed-form solution of the proximal step is a
\emph{per-voxel soft-threshold}
\begin{equation}
    \mathcal{S}_{\lambda_k \mathbf{w}_k}(u_n)
    =
    \frac{u_n}{|u_n|}
    \max\!\bigl(|u_n| - \lambda_k w_{k,n},\; 0\bigr),
    \label{eq:wprox}
\end{equation}
where $\lambda_k > 0$ is a per-layer scalar base-threshold and
the effective threshold at voxel $n$ is $\tau_{k,n} = \lambda_k w_{k,n}$.
Substituting~\eqref{eq:wprox} into~\eqref{eq:lista} yields the
W-LISTA update
\begin{equation}
    \mathbf{z}^{(k)}
    =
    \mathcal{S}_{\lambda_k\mathbf{w}_k}
    \!\left(
    \mathbf{z}^{(k-1)}
    -
    \mu_k\,
    \mathbf{T}^{\mathrm{H}}\!\left(\mathbf{T}\mathbf{z}^{(k-1)} - \mathbf{y}\right)
    \right),
    \label{eq:wlista}
\end{equation}
with the same MF warm start
$\mathbf{z}^{(0)} = \mathbf{T}^{\mathrm{H}}\mathbf{y}$ as LISTA.

\subsection{Rank-1 factorized separable weights parametrization}
Storing $N$ independent weights per layer would introduce
$K\!\times\! N$ parameters ($K\!\times\! N \!\sim\! 10^7$), which corresponds to an extreme over-parameterization.
To obtain an efficient representation, we exploit the
\emph{separability} of the 3D voxel grid along the three Cartesian axes and
express each weight field as a rank-1 outer product of three 1-D vectors according to
\begin{equation}
    w_{k,n}(x_n, y_n, z_n)
    =
    w_{k,n}^{(x)}(x_n)\cdot
    w_{k,n}^{(y)}(y_n)\cdot
    w_{k,n}^{(z)}(z_n),
    \label{eq:factorized}
\end{equation}
where
\begin{align}
\mathbf{w}_k^{(x)} &= [w_{k,1}^{(x)}(x_1),...,w_{k,N_{x}}^{(x)}(x_{N_{x}})]\in \mathbb{R}_{+}^{N_x} \\
\mathbf{w}_k^{(y)} &=  [w_{k,1}^{(y)}(y_1),...,w_{k,N_{y}}^{(y)}(y_{N_{y}})] \in \mathbb{R}_{+}^{N_y}\\
\mathbf{w}_k^{(z)} &= [w_{k,1}^{(z)}(z_1),...,w_{k,N_{z}}^{(z)}(z_{N_{z}})] \in \mathbb{R}_{+}^{N_z}
\label{factorization}
\end{align}
are the learnable 1-D profiles along the $x$-, $y$-, and $z$-axis,
respectively.
The full $N$-dimensional vector $\mathbf{w}_k$ is obtained by
flattening the $N_x \!\times\! N_y \!\times\! N_z$ tensor 
\begin{equation}
\mathbf{w}_k\approx\mathbf{w}_k^{(x)} \otimes \mathbf{w}_k^{(y)} \otimes \mathbf{w}_k^{(z)}
\end{equation}
in the same row-major order used to index the voxel positions in
$\mathbf{T}$.

The proposed parametrization~\eqref{eq:factorized} encodes the prior belief that
the spatial weight field is well-approximated by a separable function, namely that the probability of target presence can be factored as independent
marginals along each axis.
This assumption is reasonable for indoor scenes with known floor-plan geometry, where the target occupancy region forms a roughly axis-aligned sub-volume. The main limitation of the approach is that it might admit phantom responses at off-target depths. 
\textcolor{black}{Moreover, the spatial prior is expected to be reliable within the region spanned by the training positions and to over-penalize targets located in unexplored parts of the volume. The separable parametrization partly mitigates this effect, as the learned prior
extends to combinations of the per-axis coordinates observed during training,
but it does not extrapolate beyond their range. In the experiments of Sections~\ref{sec:simulations} and \ref{sec:indoor}, the validation positions lie within this region; assessing the prior on
out-of-range positions is left for future work.}

The full W-LISTA trainable parameter set is 
\begin{equation}
    \bm{\theta}
    =
    \bigl\{
    \mu_k,\;
    \lambda_k,\;
    \mathbf{w}_k^{(x)},\;
    \mathbf{w}_k^{(y)},\;
    \mathbf{w}_k^{(z)}
    \bigr\}_{k=1}^{K}.
    \label{eq:wlista_params}
\end{equation}
The total per-layer parameter count is $|\bm{\theta}|=K\bigl(2 + N_x + N_y + N_z\bigr)$,
which is several orders of magnitude smaller than the
full per-voxel parameterization.

\textcolor{black}{The 1-D weight profiles are parametrized in element-wise logarithmic scale,
$\mathbf{w}_k^{(\cdot)} = \exp\bigl(\tilde{\mathbf{w}}_k^{(\cdot)}\bigr)$, so the
log-weights $\tilde{\mathbf{w}}_k^{(\cdot)}$ are the trainable parameters, initialized as
$\tilde{\mathbf{w}}_k^{(\cdot)} = \mathbf{0}$, i.e., $\mathbf{w}_k = \mathbf{1}$, for all axes and layers.} The training objective is the same MSE loss as
in~\eqref{eq:lista_loss}, and the same Adam optimizer
with gradient clipping is employed.

\subsection{W-LISTA training and implementation considerations}
\label{sec:wlista_exp}

This section details the imaging geometry, the training set, and
the W-LISTA hyper-parameters used in all experiments reported in the
remainder of the paper, with main values summarized in
Table~\ref{tab:geom} and Table~\ref{tab:arch}. 

\textit{Imaging geometry.}
As further described in Section~\ref{sec:simulations}, the imaging domain spans $x \in [0,\,5]$\,m, $y \in [0,\,2.5]$\,m,
$z \in [0.3,\,2.3]$\,m, discretized on a uniform grid of
$N_x \!\times\! N_y \!\times\! N_z = 161 \!\times\! 81 \!\times\! 65$
voxels with isotropic spacing $\Delta = 0.031$\,m\,$\approx \lambda/4$
at the operating frequency $f_0 = 2.45$\,GHz. 

\textit{W-LISTA architecture and initialization.}
The network depth, namely the number of W-LISTA layers, is set to the number
of ISTA iterations, $N_\mathrm{iter}=K = 10$, beyond which the classical
solver \textcolor{black}{was observed to stop improving on the same data. The
depth is thus inherited from the convergence of the solver being unrolled,
rather than tuned on the learned models.} The per-layer step sizes and
base-thresholds are initialized to
$\mu_k^{(0)} = 1/L \approx 8.76 \!\times\! 10^{-5}$ and
$\lambda_k^{(0)} = 1 \!\times\! 10^{-4}$.
The factorized weight profiles are initialized so that
$\mathbf{w}_k = \mathbf{1}$ and W-LISTA reduces exactly to plain
LISTA at the start of training. Each forward pass uses the
MF warm start \eqref{eq:mf}.
The total trainable parameter
count is $|\bm{\theta}| = K(2 + N_x + N_y + N_z) = 3090$.

\textit{Optimization.}
Training minimizes the MSE loss of \eqref{eq:lista_loss}
over $P$ of scenes and for $E=30$ epochs using the Adam optimizer
configured with two parameter groups: a base learning rate
$\eta = 5 \!\times\! 10^{-2}$ for the scalar parameters
$\{\tilde{\mu}_k,\tilde{\lambda}_k\}$ and a $10\!\times$ larger
learning rate $\eta_w = 5 \!\times\! 10^{-1}$ for the spatial weight
log-vectors
$\tilde{\mathbf{w}}_k^{(x)},\tilde{\mathbf{w}}_k^{(y)},\tilde{\mathbf{w}}_k^{(z)}$.
The asymmetric schedule compensates for the smaller per-element
gradient produced by the separable parameterization, where entry of $\mathbf{w}_k^{(\cdot)}$ contributes simultaneously to a 2-D slice of voxels.
\textcolor{black}{Gradients are clipped at unit norm before each update, and the log-weights are
then bounded to $|\tilde{\mathbf{w}}_k^{(\cdot)}| \le 4$ to prevent
exponential increase of the spatial weights.
Similarly as before, the holographic operator $\mathbf{T}$ is evaluated on-the-fly via batched products.}

\begin{table}[t]
	\centering
	\caption{Imaging geometry and datasets used for validation: full-wave FEKO simulations and in-lab measurements. \textcolor{black}{Each measured position was acquired in 2 sessions; training uses the session 1, validation the session 2.}}
	\label{tab:geom}
	\setlength{\tabcolsep}{4pt}
	\small
	\begin{tabular}{@{}l >{\centering\arraybackslash}p{2.15cm} >{\centering\arraybackslash}p{2.15cm}@{}}
		\toprule
		\textbf{Parameter} & \multicolumn{2}{c}{\textbf{Value}} \\
		\midrule
		\multicolumn{3}{@{}l}{\emph{Imaging geometry (common)}} \\
		Operating frequency $f_0$        & \multicolumn{2}{c}{$2.45$\,GHz} \\
		Wavelength $\lambda$             & \multicolumn{2}{c}{$0.121$\,m} \\
		Imaging volume $x,y,z$           & \multicolumn{2}{c}{$5\!\times\!2.5\!\times\!2$\,m$^3$} \\
		Voxel spacing $\Delta$           & \multicolumn{2}{c}{$0.031$\,m\,$(\approx\!\lambda/4)$} \\
		Grid $N_x\!\times\!N_y\!\times\!N_z$ & \multicolumn{2}{c}{$161\!\times\!81\!\times\!65$} \\
		Total voxels $N$                 & \multicolumn{2}{c}{$847\,665$} \\
		Array elements $M$               & \multicolumn{2}{c}{$12\,960$ ($162\!\times\!80$)} \\
		Lipschitz constant $L$           & \multicolumn{2}{c}{$1.141\!\times\!10^{4}$} \\
		\midrule
		\multicolumn{3}{@{}l}{\emph{Training and validation set}} \\
		                                 & \textbf{Synthetic} & \textbf{Real} \\
		Source                           & FEKO full-wave & in-lab meas. \\
		\textcolor{black}{Positions}       & \textcolor{black}{$11$} & \textcolor{black}{$8$} \\
		Training meas. $P$               & \textcolor{black}{$6$} & \textcolor{black}{$8$ (session $1$)} \\
		Validation meas.                 & \textcolor{black}{$5$} & \textcolor{black}{$5$
(session $2$)} \\
		Supervisory contrast $\Delta\varepsilon$ & $1.53$ & $1.53$ \\
		Active voxels/scene              & $\sim\!6\,300$ ($\sim\!0.74\%$) & $\sim\!6\,300$ ($\sim\!0.74\%$) \\
		\bottomrule
	\end{tabular}
\end{table}

\begin{table}[t]
	\centering
	\caption{W-LISTA and LoRaW-LISTA:
	architecture, initialization and training hyper-parameters. The low-rank operator correction $\{\mathbf{U},\mathbf{V}\}$ refers to LoRaW-LISTA only.}
	\label{tab:arch}
	\setlength{\tabcolsep}{3pt}
	\renewcommand{\arraystretch}{1.15}
	\footnotesize
	\resizebox{\columnwidth}{!}{%
	\begin{tabular}{@{}p{3.15cm} l l@{}}
		\toprule
		\textbf{Parameter} & \textbf{W-LISTA} & \textbf{LoRaW-LISTA} \\
		\midrule
		\multicolumn{3}{@{}l}{\emph{Architecture}} \\
		Unrolled layers $K$                                                        & $10$ & $10$ \\
		Forward operator                                                           & $\mathbf{T}$ (fixed) & $\mathbf{T}_{\mathrm{eff}}=(\mathbf{I}_M+\mathbf{U}\mathbf{V}^\mathrm{H})\,\mathbf{T}$ \\
		Rank $r$                                                                   & --   & $8$ \\
		Warm start $\mathbf{z}^{(0)}$                                              & $\mathbf{T}^{\mathrm{H}}\mathbf{y}$ & $\mathbf{T}_{\mathrm{eff}}^{\mathrm{H}}\mathbf{y}$ \\
		Trainable params                                          & $3090$     & $\approx\!4.1\!\times\!10^{5}$ ($r=8$) \\
		\midrule
		\multicolumn{3}{@{}l}{\emph{Initialization}} \\
		Step size $\mu_k^{(0)}$                                                     & $1/L\approx10^{-4}$ & $1/L\approx10^{-4}$ \\
		Base threshold $\lambda_k^{(0)}$                                           & $1\!\times\!10^{-4}$ & $1\!\times\!10^{-4}$ \\
		Spatial log-weights $\tilde{\mathbf{w}}_k^{(\cdot)}$                                   & $\mathbf{0}$ ($\mathbf{w}_k^{(\cdot)}=\mathbf{1}$) & $\mathbf{0}$ ($\mathbf{w}_k^{(\cdot)}=\mathbf{1}$) \\
		Low-rank factor $\mathbf{U}$, $\mathbf{V}$                                               & --   & $\mathbf{0}$, $\mathcal{N}(0,10^{-2})$ \\
		\midrule
		\multicolumn{3}{@{}l}{\emph{Training} (Optimizer Adam)} \\
		Epochs                                                                     & $30$ & $30$ \\
		LR for $\mu_k,\lambda_k$ ($\eta$)                                          & $5\!\times\!10^{-2}$ & $5\!\times\!10^{-2}$ \\
		LR for $\tilde{\mathbf{w}}_k^{(\cdot)}$ ($\eta_w$)                                  & $2.5\text{--}5\!\times\!10^{-1}$ & $2.5\text{--}5\!\times\!10^{-1}$ \\
		LR for $\mathbf{U},\mathbf{V}$ ($\eta_{\mathrm{lr}}$)                       & --   & $1\!\times\!10^{-4}$ \\
		Loss                                                                       & MSE~\eqref{eq:lista_loss} & Eq.~\eqref{eq:lrwlista_loss} \\
		Reconstruction weight $\alpha_z$                                      & $1$  & $1$ \\
		Data-consistency weight $\beta_d$                                          & --   & $1\!\times\!10^{-4}$ \\
		Regularization\ weight $\gamma$                                             & --   & $0.1$ \\
		\bottomrule
	\end{tabular}%
	}
\end{table}

\section{Low-Rank Adaptation of the Holographic Operator}
\label{sec:lrwlista}

The W-LISTA architecture of Section~\ref{sec:wlista} keeps the
holographic operator $\mathbf{T}$ fixed and learns only the
regularization environment, i.e., the per-layer step sizes, base
thresholds and factorized spatial weights. This keeps the network
compact and fast to train, and is well suited whenever the linearized
forward model $\mathbf{y} = \mathbf{T}\mathbf{z}$ is an accurate
description of the physics.

In this section we introduce a Low-Rank operator correction for the
Weighted LISTA (LoRaW-LISTA) that additionally learns a low-rank
adaptation of the holographic operator $\mathbf{T}$, which might be
inaccurate for echoic environments and, for certain types of objects, suffer from inconsistencies due to the Born
approximation. LoRaW-LISTA
augments W-LISTA with a supervised, low-rank correction of $\mathbf{T}$
while preserving its physical interpretation. The
trade-off between the two variants is assessed experimentally in the
following sections.

\subsection{Low-Rank Weighted LISTA (LoRaW-LISTA)}

Following the same principle as low-rank adaptation (LoRA) of large
pre-trained models~\cite{hu2021lora}, we keep the physics-based Born
operator $\mathbf{T}$ frozen and learn only a low-rank correction term on top
of it. We define $\mathbf{T}_{\mathrm{eff}}$ as a low-rank $r\!\ll\!M$ perturbation of the holographic operator $\mathbf{T}$ under Born assumptions via
\begin{equation}
    \mathbf{T}_{\mathrm{eff}}
    =
    \bigl(\mathbf{I}_M + \mathbf{U}\mathbf{V}^\mathrm{H}\bigr)\,\mathbf{T},
    \qquad
    \mathbf{U},\mathbf{V}\in\mathbb{C}^{M\times r},
    \label{eq:teff}
\end{equation}
where $\mathbf{U}\mathbf{V}^\mathrm{H}$ is a rank$-r$ perturbation acting on the $M$ receiver samples: it corrects the Born-predicted fields $\mathbf{T}\mathbf{z}$ towards those observed in the training (target) data, thus compensating the model mismatch of $\mathbf{T}$.
The LoRaW-LISTA update is obtained by
substituting $\mathbf{T}\!\to\!\mathbf{T}_{\mathrm{eff}}$ in the
W-LISTA recursion~\eqref{eq:wlista}, which leads to
\begin{equation}
    \mathbf{z}^{(k)}
    =
    \mathcal{S}_{\lambda_k\mathbf{w}_k}
    \!\left(
    \mathbf{z}^{(k-1)}
    -
    \mu_k\,
    \mathbf{T}_{\mathrm{eff}}^\mathrm{H}\!\left(\mathbf{T}_{\mathrm{eff}}\mathbf{z}^{(k-1)} - \mathbf{y}\right)
    \right),
    \label{eq:lrwlista}
\end{equation}
with warm start
$\mathbf{z}^{(0)} = \mathbf{T}_{\mathrm{eff}}^\mathrm{H}\mathbf{y}$. Similarly as for~\eqref{eq:wlista}, replacing
$\mathbf{T}$ by $\mathbf{T}_{\mathrm{eff}}$ also requires the
corresponding $\mathbf{T}_{\mathrm{eff}}^{\mathrm{H}}$, which follows from~\eqref{eq:teff} as $\mathbf{T}_{\mathrm{eff}}^\mathrm{H} \mathbf{r}
    =
    \mathbf{T}^\mathrm{H}\!\bigl[\mathbf{r} + \mathbf{V}(\mathbf{U}^\mathrm{H}\mathbf{r})\bigr]$ and $\mathbf{r}=\mathbf{T}_{\mathrm{eff}}\mathbf{z}^{(k-1)} - \mathbf{y}$.
Notice that the factors $\mathbf{U},\mathbf{V}$ are \emph{shared} across all $K$
layers and are
initialized as $\mathbf{U}=\mathbf{0}$ (and $\mathbf{V}$ random),
so that ${\mathbf{T}_{\mathrm{eff}}=\mathbf{T}}$.

\subsection{Reconstruction and data-consistency loss}

The reconstruction  loss of~\eqref{eq:lista_loss} is invariant to the
phase of $\mathbf{z}$ and therefore provides no gradient able to
train a phase correction such as $\mathbf{U}\mathbf{V}^\mathrm{H}$. The objective is thus augmented with a \textit{data-consistency term},
which supplies the phase-sensitive signal needed to fit
$\Delta\mathbf{T}=\mathbf{U}\mathbf{V}^\mathrm{H}$. The proposed loss function 
\begin{equation}
\begin{aligned}
    \mathcal{L}(\bm{\theta})
    =\;&
    \alpha_z\,\frac{1}{P}\sum_{i=1}^{P}
    \bigl\| |\hat{\mathbf{z}}_i| - |\mathbf{z}_i^*| \bigr\|_2^2
    \\[2pt]
    +\;&
    \beta_d\,\frac{1}{P}\sum_{i=1}^{P}
    \frac{\bigl\|\mathbf{T}_{\mathrm{eff}}\hat{\mathbf{z}}_i - \mathbf{y}_i\bigr\|_2^2}
         {\|\mathbf{y}_i\|_2^2}
    \;+\;
    \gamma\,\bigl\|\mathbf{U}\mathbf{V}^\mathrm{H}\bigr\|_F^2 .
\end{aligned}
\label{eq:lrwlista_loss}
\end{equation}
also includes a Frobenius regularizer.
The first term is the reconstruction loss~\eqref{eq:lista_loss}
inherited from W-LISTA, the second is a data-consistency
term that trains $\mathbf{U},\mathbf{V}$, and the third regularizes the low-rank correction. 
The normalized data-consistency term supervises the
low-rank correction through the residual
$\|\mathbf{T}_{\mathrm{eff}}\hat{\mathbf{z}}-\mathbf{y}\|_2^2/
\|\mathbf{y}\|_2^2$. The Frobenius regularizer
$\|\mathbf{U}\mathbf{V}^\mathrm{H}\|_F^2$ discourages corrections larger than necessary. The LoRaW-LISTA parameters are now $\bm{\theta}\cup\{\mathbf{U},\mathbf{V}\}$.


\begin{figure}[t]
    \centering
    \includegraphics[width=\linewidth]{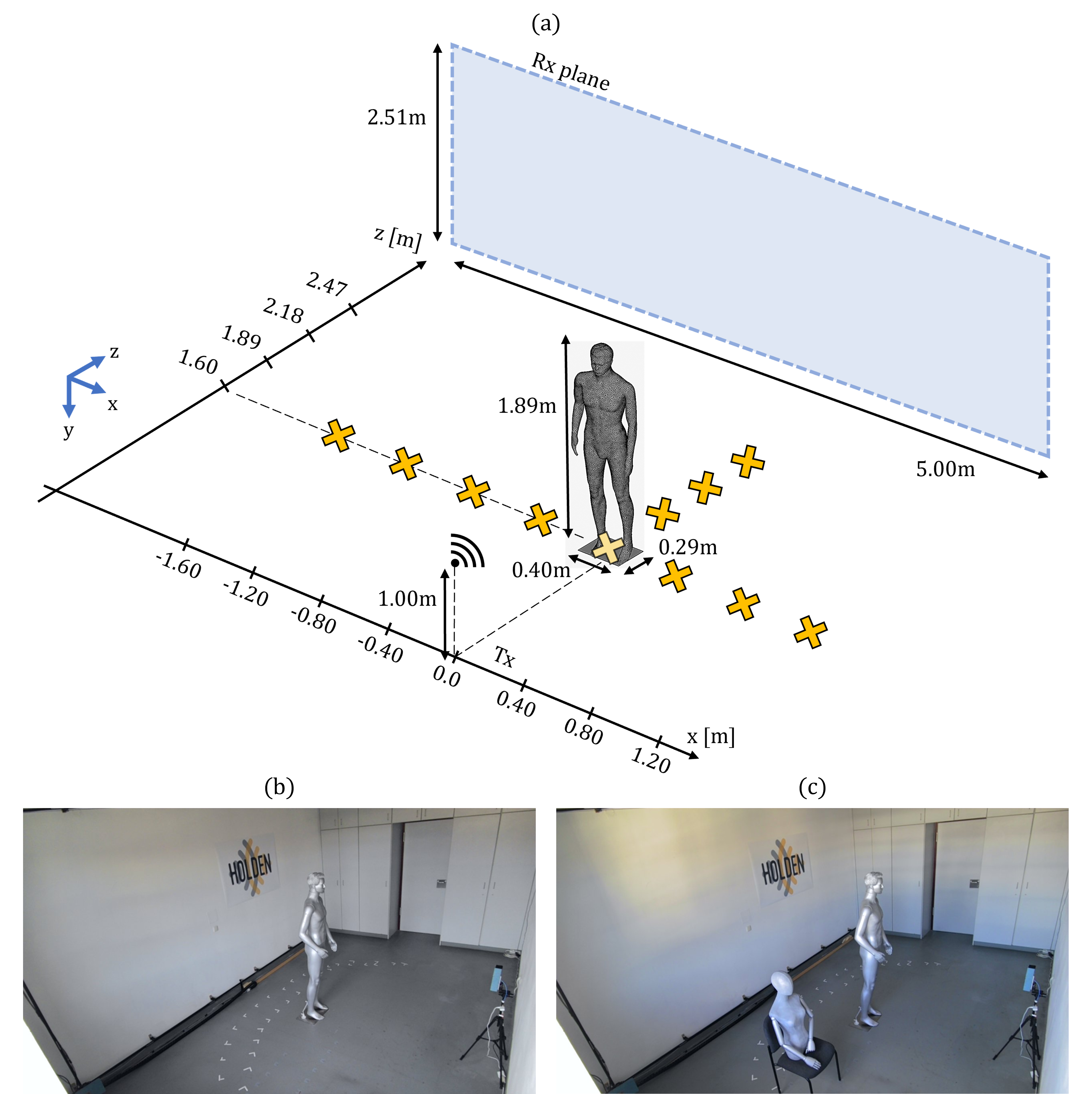}
    \caption{Indoor measurement scenario. (a) Acquisition geometry in the Tx-centered reference frame (axis triad shown top-left; Tx at $1.00$\,m height): the $11$ target positions (crosses) listed in Table~\ref{tab:positions} over the $5.00\!\times\!2.51$\,$\mathrm{m}^2$ vertical Rx
plane. (b) Single- and (c) two-phantom experimental configuration.}
    \label{fig:meas_scenario}
\end{figure}

\subsection{LoRaW-LISTA training and computational cost}
\label{sec:lrwlista_exp}

LoRaW-LISTA inherits the imaging geometry, the W-LISTA backbone and
the W-LISTA training procedure of Section~\ref{sec:wlista_exp}, and
extends them with the low-rank operator correction. Architecture,
initialization and training hyper-parameters are summarized in
Table~\ref{tab:arch}. Two-stage algorithm implementation is shown in Algorithm \ref{alg:wfirst}.

\textit{Rank and initialization.}
The rank $r$ trades expressiveness of the operator correction against
parameter count, training time and risk of over-fitting. Setting $r=8$ is found as sufficient to capture the dominant residual scattering unexplained by the linearized model of \eqref{born}, while keeping the trainable parameter budget limited as below one order of magnitude above the receiver count $M$. The low-rank
factors are initialized asymmetrically~\cite{hu2021lora} (see Algorithm~\ref{alg:wfirst}), so that the initial
correction $\mathbf{U}\mathbf{V}^\mathrm{H}=\mathbf{0}$ and LoRaW-LISTA
coincides with W-LISTA at the start of training.

\textit{Optimization.}
The reconstruction ($\alpha_z$), data-consistency ($\beta_d$) and regularization ($\gamma$) loss weights are chosen in Table \ref{tab:arch} to guarantee stability. Adam is configured with three parameters: the W-LISTA learning rates $\eta$ and $\eta_w$,
unchanged from Section~\ref{sec:wlista_exp}, and an additional
learning rate $\eta_{\mathrm{lr}}=1\!\times\! 10^{-2}$ used for the
low-rank factors. $\eta_{\mathrm{lr}}$ should be kept small since $\mathbf{U}\mathbf{V}^\mathrm{H}$ enters multiplicatively into the
forward operator $\mathbf{T}_{\mathrm{eff}}$, therefore, even small
updates propagate through every layer of the unrolled network.

\textit{Two-stage training.}
The spatial $\ell_1$ regularization of W-LISTA and the low-rank adaptation of the
holographic operator are learned sequentially following the Algorithm~\ref{alg:wfirst}. In a first
warm-up stage (epochs $1,\dots,E_w$), the low-rank factors are frozen at
$\mathbf{U}\mathbf{V}^{\mathrm H}=\mathbf{0}$, so that $\mathbf{T}_{\mathrm{eff}}=\mathbf{T}$
and only the W-LISTA parameters are
trained. In a second stage
(epochs $E_w{+}1,\dots,E$), the low-rank factors are unfrozen and all parameters are
optimized jointly under the composite loss~\eqref{eq:lrwlista_loss}, which activates the
data-consistency term supervising the operator correction. Stabilizing the spatial weights
before activating the data-consistency term (so implementing W-LISTA as warm start) prevents the weights and the low-rank
correction from interacting uncontrollably destabilizing the training; in our experiments $E_w=6$ warm-up epochs out of $E=30$ are used. Finally, note that performing the two phases in the reverse order (low-rank adaptation first, and spatial regularization second) yields noticeably poorer reconstructions.

\begin{algorithm}[t]
\caption{LoRaW-LISTA implementation: spatial $\ell_1$ regularization, followed by low-rank adaptation of  $\mathbf{T}$}
\label{alg:wfirst}
\begin{algorithmic}[1]
\Require training $\{(\mathbf{y}_i,\mathbf{z}_i)\}$, operator $\mathbf{T}$, layers $K$, warm-up epochs $E_w<E$, total epochs $E$, learning rates $\eta,\eta_w,\eta_{\mathrm{lr}}$
\State Initialize W-LISTA parameters $\bm{\theta}$ in \eqref{eq:wlista_params}
\State Initialize low-rank factors $\mathbf{U}=\mathbf{0}$, $\mathbf{V}\sim\mathcal{N}(0,10^{-2})$ 
\Statex \emph{Stage 1 -- warm-up (W-LISTA spatial $\ell_1$ regularization)}
\For{epoch $=1$ to $E_w$}
  \State freeze $\{\mathbf{U},\mathbf{V}\}$; \quad $\mathbf{T}_{\mathrm{eff}}\leftarrow\mathbf{T}$
  \State $\hat{\mathbf{z}}_i\leftarrow\text{W-LISTA}_{\bm{\theta}}(\mathbf{y}_i,\mathbf{T}_{\mathrm{eff}})$ for all $i=1,...,P$
  \State update $\bm{\theta}$ (Adam, rates $\eta,\eta_w$) on the MSE loss~\eqref{eq:lista_loss}
\EndFor
\Statex \emph{Stage 2 -- joint low-rank adaptation of $\mathbf{T}$}\\ Update parameters $\bm{\theta}\leftarrow\bm{\theta}\cup\{\mathbf{U},\mathbf{V}\}$ 
\For{epoch $=E_w{+}1$ to $E$}
  \State unfreeze $\{\mathbf{U},\mathbf{V}\}$; \quad $\mathbf{T}_{\mathrm{eff}}\leftarrow(\mathbf{I}_M+\mathbf{U}\mathbf{V}^{\mathrm H})\,\mathbf{T}$
  \State $\hat{\mathbf{z}}_i\leftarrow\text{W-LISTA}_{\bm{\theta}}(\mathbf{y}_i,\mathbf{T}_{\mathrm{eff}})$ for all $i$
  \State update $\{\bm{\theta}\}$ on  loss $\mathcal{L}(\bm{\theta})$~\eqref{eq:lrwlista_loss}
\EndFor
\end{algorithmic}
\end{algorithm}

\textit{Computational cost.}
The low-rank correction adds two $M\!\times\!r$ products per
matrix-vector evaluation, negligible against the on-the-fly
evaluation of the physical kernel $\mathbf{T}$. The parameter budget
grows from the W-LISTA value $K(2+N_x+N_y+N_z)=3090$ to
$|\bm{\theta}| = K(2+N_x+N_y+N_z) + 4 M r$, where $4Mr$ counts the
real and imaginary parts of
$\mathbf{U},\mathbf{V}\in\mathbb{C}^{M\times r}$. For the considered deployment with $M=12\,960$ array elements (see Section~\ref{sec:simulations}) and $r=8$, this amounts to
$4 M r \approx 4.1\!\times\! 10^{5}$ trainable parameters, roughly two
orders of magnitude more than W-LISTA. LoRaW-LISTA is thus best motivated when a measurable
forward-model mismatch is present.

\begin{figure*}[t]
    \centering
    \includegraphics[scale=0.9]{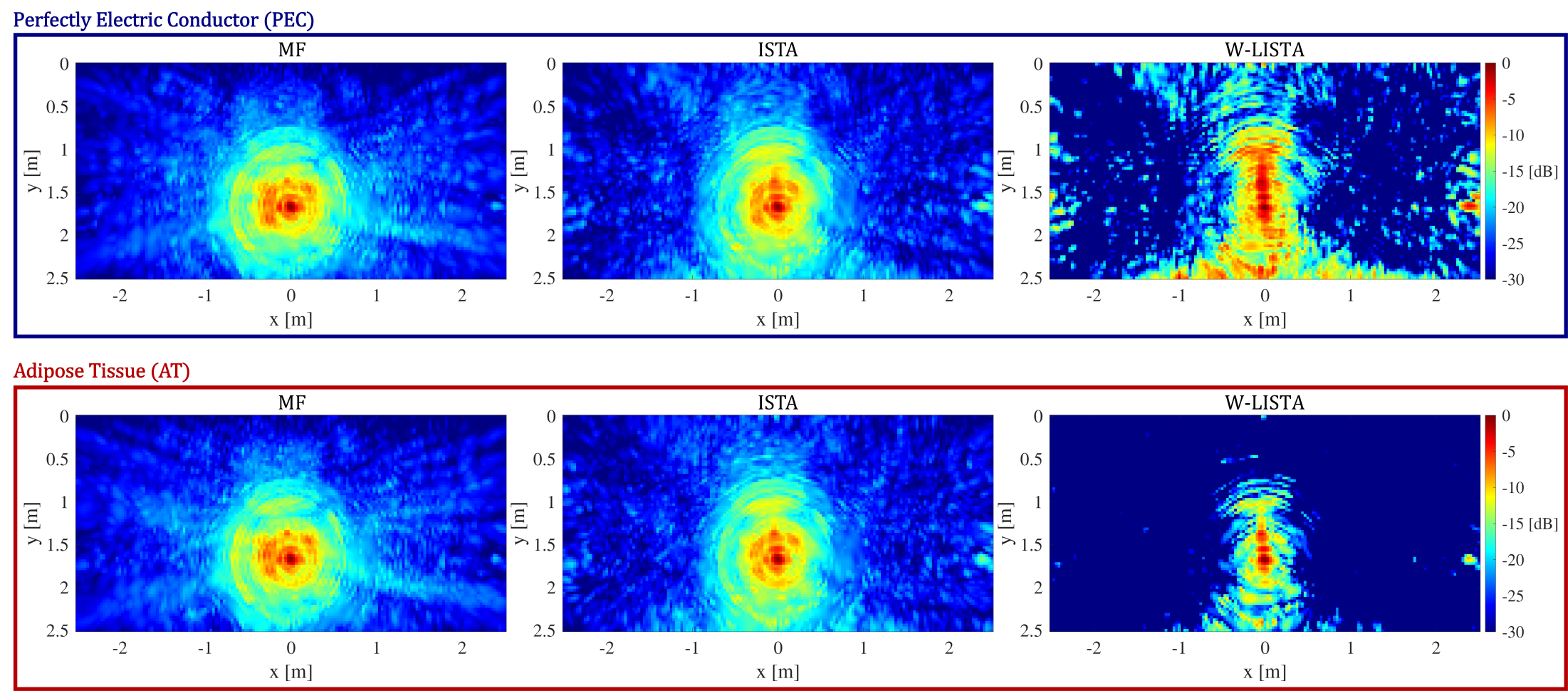}
    \caption{Synthetic holographic reconstruction of a human body located at the centre of the room ($q=9$, as in Fig.~\ref{fig:meas_scenario}a), for two body-tissue models, from top to bottom: Perfectly Electric Conductor (PEC) and Adipose Tissue (AT). For each tissue, the MP projection of the reconstructed occupancy obtained with matched-filter (MF) back-projection, ISTA and the proposed W-LISTA.}
    \label{fig:synt_cfr}
\end{figure*}

\begin{figure}[t]
    \centering
    \includegraphics[trim={2cm 1cm 10cm 6cm}, clip, width=\linewidth]{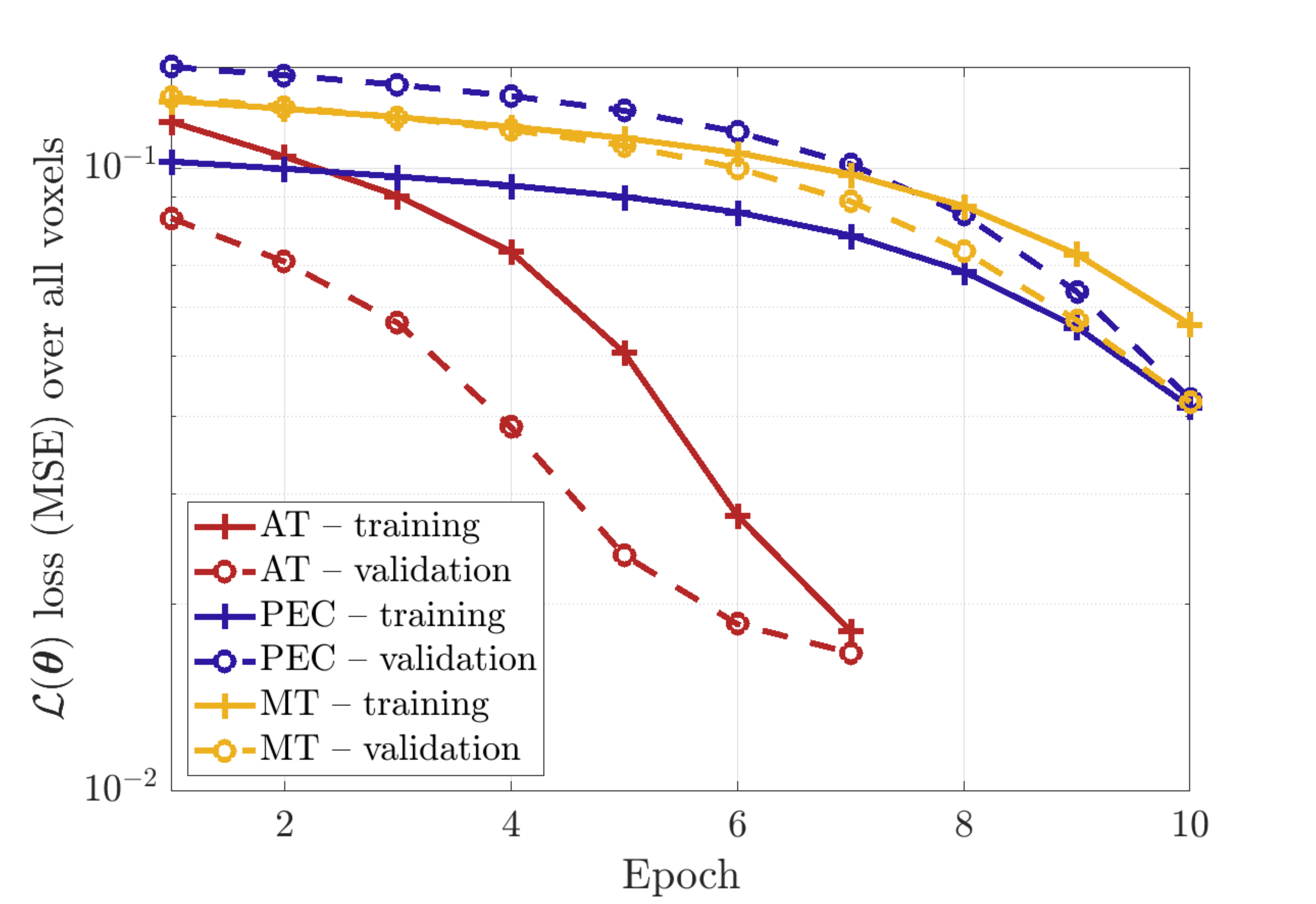}
    \caption{Training and validation reconstruction loss of W-LISTA (evaluated over all the 3D voxel space) versus training epochs, for the three body-tissue models: PEC, MT and AT.}
    \label{fig:loss_at_vs_pec}
\end{figure}

\section{EM simulations and results}
\label{sec:simulations}

This section presents the validation of the unrolling tools on synthetic EM data, generated through full-wave simulation of the RF propagation in the target environment. The full-wave model reproduces the indoor measurement scenario of Section~\ref{sec:indoor} using the method of moments, with a mesh size of $\lambda/10$, where $\lambda$ denotes the free-space wavelength at the operating frequency of 2.45\,GHz. The imaging target is a human body, whose spatial occupancy and shape are reconstructed from the scattered field via RF holography. To probe the holographic reconstruction of the body silhouette across the full range of scattering behaviours, the human body is modelled under three dielectric representations. 
The spatially-varying $\ell_1$ regularization of W-LISTA is benchmarked against the state-of-the-art MF and ISTA baselines in \eqref{eq:mf} and \eqref{l1_reg}, respectively. The sparse regularized ISTA solution is obtained with ${\alpha=10^{-4}}$, in line with W-LISTA. The low-rank operator adaptation (LoRaW-LISTA) is assessed on real measurements in the case study of Section~\ref{sec:indoor}. The simulation scenarios and body models are detailed in Section~\ref{sec:simulations}-A, and the reconstruction results and loss analysis are reported in Section~\ref{sec:simulations}-B. Part of the code and the dataset are made available online\footnote{Reference code and dataset: \url{https://github.com/labRadioVision/unrolled_holographic_imaging}}.




\subsection{Human body representation}

\begin{table*}[t]
\centering
\caption{Target positions in a Tx-centered frame (Fig.~\ref{fig:meas_scenario}):
$x_q$ lateral, $z_q$ depth, at fixed floor height.}
\label{tab:positions}
\begin{tabular}{@{}l ccccccccccc@{}}
\toprule
\textbf{Position} $q$ & 1 & 2 & 3 & 4 & 5 & 6 & 7 & 8 & 9 & 10 & 11 \\
\midrule
$x_q$ [m] & $-0.40$ & $-0.80$ & $-1.20$ & $-1.60$ & $0.40$ & $0.80$ & $1.20$ & $0.00$ & $0.00$ & $0.00$ & $0.00$ \\
$z_q$ [m] & $1.60$ & $1.60$ & $1.60$ & $1.60$ & $1.60$ & $1.60$ & $1.60$ & $1.60$ & $1.89$ & $2.18$ & $2.47$ \\
\bottomrule
\end{tabular}
\end{table*}

The room geometry is illustrated in Fig.~\ref{fig:meas_scenario}. The transmitter is placed at 10\,cm from one wall, at a height of 1\,m above the floor, and centered along the wall length to ensure sufficiently uniform illumination of the room. It is modelled as a single, center-fed, vertically oriented electric dipole. The receiver array consists of $162\!\times\!80$ elements, with horizontal and vertical spacings of 0.0309\,m and 0.0314\,m, respectively; only the vertical electric field component is sampled, without explicitly modeling the patch antenna pattern. The spacing between the receiver array and the side wall is 28.3\,cm, resulting in a transmitter-to-receiver distance of approximately 3.217\,m.
Multipath propagation is accounted for via the image-source formulation described in~\cite[pp.~94--99]{Jin.2015}, whose validity for this environment was verified in~\cite{eucap26}. Six effective complex reflection coefficients, optimized in~\cite{eucap26}, are adopted:
\begin{align}
\Gamma_{1} &= 0.19\angle 95^\circ \quad (\text{right wall}), &
\Gamma_{2} &= 0.15\angle 55^\circ \quad (\text{left wall}), \nonumber\\
\Gamma_{3} &= 0.11\angle 0^\circ \quad (\text{ceiling}), &
\Gamma_{4} &= 0.17\angle 17^\circ \quad (\text{ground}), \nonumber\\
\Gamma_{5} &= 0.11\angle 243^\circ \quad (\text{back-Rx}), &
\Gamma_{6} &= 0.70\angle 287^\circ \quad (\text{back-Tx}). \nonumber
\end{align}

The human body is represented by a CAD phantom measuring 1.89\,m in height and 51\,cm in width, placed on a metallic base plate of $40\!\times\!29$\,cm$^2$. The target is placed at the $11$ positions detailed in
Table~\ref{tab:positions}, comprising $8$ lateral displacements at
fixed depth and $3$ depth displacements at fixed lateral position. \textcolor{black}{Of the $11$ positions, $6$ are used for training
($q=2,4,5,7,10,11$) and the remaining $5$ for validation ($q=1,3,6,8,9$).} At all positions, the phantom faces the wall behind the transmitter.


Three material configurations are investigated: a perfectly electric conducting (PEC) body, a Muscle-Tissue (MT) body, and an Adipose-Tissue (AT) body. The lossy tissues are modelled as a homogeneous layer with thickness equal to four skin depths, ensuring sufficient field attenuation without requiring a thicker structure. MT is characterized at 2.45\,GHz by a relative permittivity $\varepsilon_r=59.77$, a dielectric loss tangent $\tan\delta=0.242$, and a mass density $\rho=1040\,\mathrm{kg/m^3}$; AT by $\varepsilon_r\approx5.3$, $\tan\delta\approx0.15$, and $\rho\approx910\,\mathrm{kg/m^3}$.

\subsection{Results and reconstruction loss analysis}
Fig.~\ref{fig:synt_cfr} shows a representative reconstruction of a human body in the center of the room ($q=9$), for the PEC (top) and AT (bottom) models. For each, the occupancy from MF back-projection, ISTA and W-LISTA is compared against the ground-truth silhouette (Fig.~\ref{fig:meas_scenario}a). MF yields a diffuse, low-resolution map dominated by the aperture point-spread function; ISTA sharpens it through $\ell_1$ regularization, but leaves large errors on the occupied voxels. The spatially-varying $\ell_1$ regularization of W-LISTA produces a cleaner, better-localized map, concentrating the reflectivity onto the body support. The improvement is clearer for AT, for which the first-order Born operator (see Section~\ref{sec:rf_holography}) is more accurate; when the permittivity is high enough to violate Born (PEC), reducing the learning rate of the spatial weights $\eta_w=2.5\!\times\!10^{-1}$ is beneficial.

\begin{table}[t]
\centering
\color{black}
\caption{Reconstruction metrics per tissue and algorithm, at the same epoch.
\textsc{full}/\textsc{occ}: MSE over all voxels / over occupied voxels only.
\textsc{avg}: mean over $q \in \{1,3,6,8,9\}$ and at the central position \texttt{$q=9$}. Best cases are in bold.}
\label{tab:occ_loss}
\setlength{\tabcolsep}{1pt}
\footnotesize
\begin{tabular*}{\columnwidth}{@{\extracolsep{\fill}}ll cccc cccc@{}}
\toprule
& & \multicolumn{4}{c}{\textsc{avg}} & \multicolumn{4}{c}{$q=9$} \\
\cmidrule(lr){3-6} \cmidrule(lr){7-10}
\textbf{Tissue} & \textbf{Method} & \textsc{full} & \textsc{occ} & SCR & SSIM & \textsc{full} & \textsc{occ} & SCR & SSIM \\
\midrule
\multirow{4}{*}{AT}  & MF      & 7.5 & 13.8 & 2.67 & 0.018 & 7.5 & 44.4 & 3.70 & 0.028 \\
                     & ISTA    & \textbf{0.016} & 2.32 & 2.53 & 0.016 & \textbf{0.017} & 2.33 & 4.08 & 0.033 \\
                     & LISTA   & 0.110 & \textbf{1.52} & 1.80 & 0.007 & 0.122 & \textbf{1.12} & 2.91 & 0.015 \\
                     & W-LISTA & 0.028 & 2.07 & \textbf{3.03} & \textbf{0.155} & 0.033 & 2.03 & \textbf{5.99} & \textbf{0.208} \\
\cmidrule(lr){1-10}
\multirow{4}{*}{MT}  & MF      & 7.7 & 21.0 & 2.60 & 0.019 & 11.4 & 68.9 & 3.56 & 0.034 \\
                     & ISTA    & \textbf{0.016} & 2.33 & 2.49 & 0.018 & \textbf{0.017} & 2.33 & 3.94 & 0.040 \\
                     & LISTA   & 0.094 & \textbf{1.60} & 1.77 & 0.007 & 0.146 & \textbf{1.08} & 2.87 & 0.013 \\
                     & W-LISTA & 0.037 & 2.02 & \textbf{3.23} & \textbf{0.268} & 0.064 & 1.87 & \textbf{6.13} & \textbf{0.258} \\
\cmidrule(lr){1-10}
\multirow{4}{*}{PEC} & MF      & 6.2 & 11.1 & 2.61 & 0.017 & 6.3 & 36.0 & 3.66 & 0.028 \\
                     & ISTA    & \textbf{0.016} & 2.34 & 2.45 & 0.016 & \textbf{0.017} & 2.33 & 3.93 & 0.032 \\
                     & LISTA   & 0.085 & \textbf{1.63} & 1.70 & 0.008 & 0.093 & \textbf{1.26} & 2.69 & 0.012 \\
                     & W-LISTA & 0.029 & 2.06 & \textbf{3.67} & \textbf{0.249} & 0.038 & 2.02 & \textbf{7.61} & \textbf{0.363} \\
\bottomrule
\end{tabular*}
\end{table}

Fig.~\ref{fig:loss_at_vs_pec} reports the training and validation reconstruction loss of W-LISTA, evaluated over the whole 3D voxel space, across the training epochs and for the three different tissue models (PEC, MT, and AT). The loss attained on the AT body is consistently smaller than that of the PEC and MT bodies: high-contrast or conducting bodies induce a larger model mismatch that the fixed operator cannot fully explain. Table~\ref{tab:occ_loss} compares the average reconstruction loss, namely the MSE \eqref{eq:lista_loss} ($\texttt{FULL}$) with the loss restricted to the occupied voxels ($\texttt{OCC}$). The MSE is shown on average ($\texttt{AVG}$) w.r.t. \textcolor{black}{the
positions $q=1,3,6,8,9$ common to the three datasets} and separately for
$q=9$. \textcolor{black}{Two image-domain metrics are also reported, both
computed on the MIP images: the Signal-to-Clutter Ratio (SCR), namely the
ratio of the mean MIP level on and off the body support, and the Structural
Similarity index (SSIM)~\cite{Wang2004SSIM}.} 

On the occupied voxels, the PEC body scores best since the errors caused by
Born model mismatch fall mostly in the background rather than on the body, so
the reconstruction stays accurate on the occupied support while background
errors penalize only the full 3D volume reconstruction loss. 
\textcolor{black}{Plain LISTA attains the lowest MSE
error among all methods, however, this is due to the amplitude scale of the
reconstruction. Benefits of the W-LISTA spatial regularization approach and structured sparsity are more evident when analyzing the SCR and SSIM metrics. W-LISTA scores the highest SCR and SSIM of
all methods, whereas LISTA scores the lowest. With respect to ISTA and to plain LISTA, the SCR improves roughly by about $20\%$ and $60\%$ on the AT body, $30\%$ and $80\%$ on the MT body, and $50\%$ and
$110\%$ on the PEC one, for which the first-order Born linearization is least accurate. The image-domain metrics are further analyzed in the following.} 



\begin{figure*}[t]
    \centering
    \includegraphics[scale=0.8]{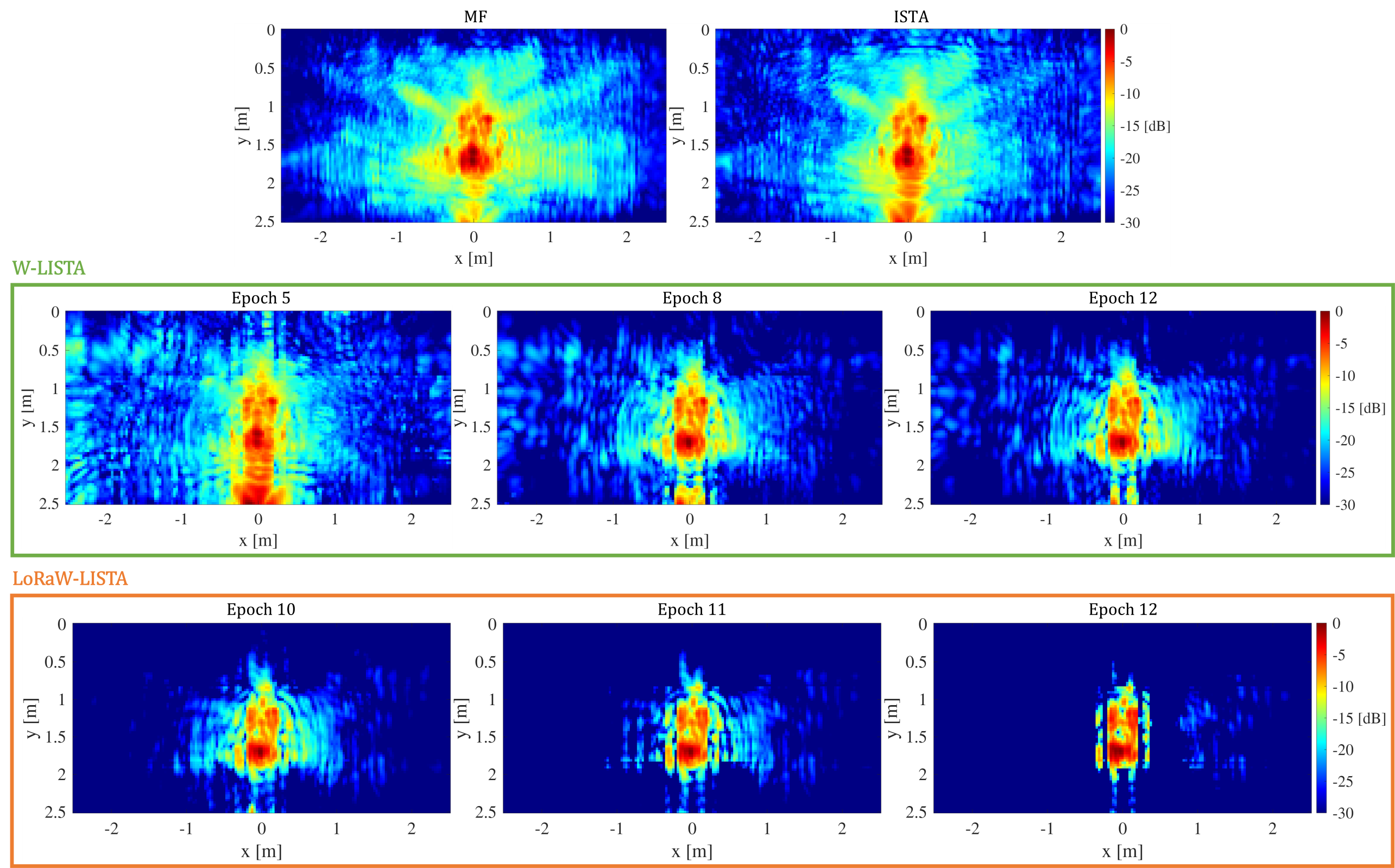}
    \caption{Reconstructed occupancy maps (MIP, dB) from real measurements
with a single phantom located at position $q=9$. Top: MF and ISTA baselines. Middle:
W-LISTA (epochs $5$, $8$, $12$). Bottom: LoRaW-LISTA (epochs $10$, $11$,
$12$). Both learned methods increasingly localize the reflectivity on
the body and suppress clutter across training, LoRaW-LISTA giving the
sharpest result.}
    \label{fig:real_cfr_epoch}
\end{figure*}

\begin{figure*}[t]
    \centering
    \includegraphics[width=\textwidth]{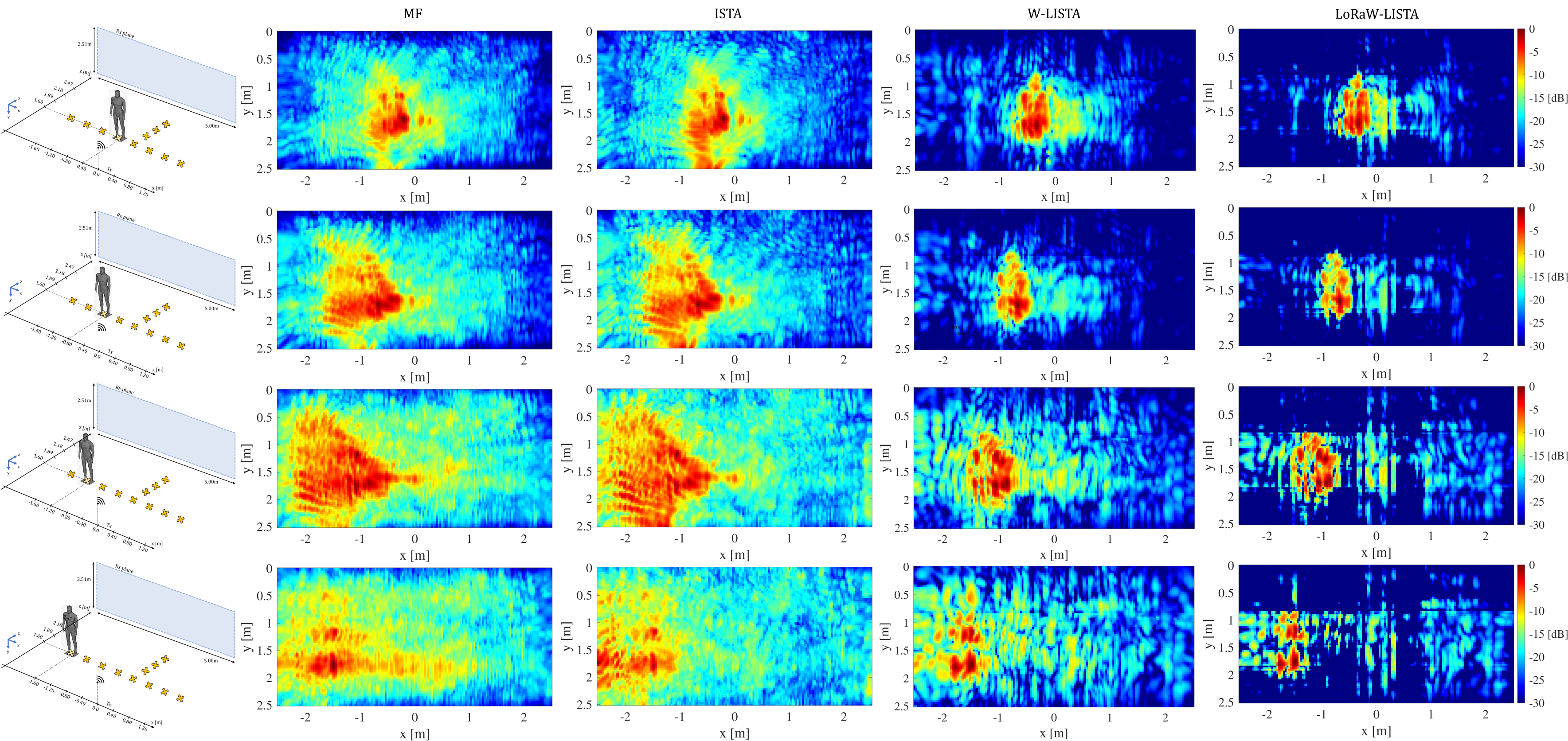}
    \caption{Reconstructed occupancy (MIP) for four distinct target
positions ($q=1,2,3,4$ in Table~\ref{tab:positions}; ground-truth
location on the left of each row), comparing MF back-projection, ISTA,
W-LISTA and LoRaW-LISTA on the real indoor measurements.}
    \label{fig:real_cfr}
\end{figure*}

\section{Indoor case study}
\label{sec:indoor}

This section validates the proposed unrolled reconstruction methods on
real indoor measurements at $2.45$\,GHz, collected with 
human-body phantoms in the laboratory environment of Fig.\,\ref{fig:meas_scenario}. The measurement setup
defines the indoor scenario and geometry that the FEKO simulations of
Section~\ref{sec:simulations} reproduce. The real data probe the robustness of the methods to calibration
errors, hardware imperfections and unmodelled multipath effects that most challenge the linearized EM model.

\subsection{Training dataset and data collection}

Reception is performed over the planar sampling grid by a single
narrowband patch antenna moved to each grid location by a two-axis
positioner, thereby synthesizing a receiving aperture consisting of $162\!\times\!80$ sample locations.
The complex transmission coefficient $S_{21}$ between Tx and Rx is
acquired around $2.45$\,GHz through a phase-stable link using a
\mbox{NanoVNA 6000-A~\cite{nanovna6000}}, providing the field samples
$b_m\propto S_{21}(\mathbf r_m)$ of~\eqref{eq:scat_meas}. For reconstruction, two types of acquisitions are
collected: an empty-room measurement, approximately providing the
incident-field reference $\mathbf b^{\mathrm{inc}}$, and target-present
measurements with the standing human phantom, whose difference yields
the calibrated scattered field
$\mathbf y = \mathbf b - \mathbf b^{\mathrm{inc}}$ of~\eqref{eq:scat_meas}.
The phantom is coated with a conductive zinc--aluminum paint, emulating
the PEC configuration of the synthetic
dataset. 

\textcolor{black}{Usable measurements are available for $8$ of the $11$ positions defined in
Table~\ref{tab:positions}, namely $q=1$--$4$ and $q=8$--$11$. Measurements were acquired in two sessions (session 1 and 2) separated by several days. The unrolled networks are trained on the first session covering  all the $8$ positions, and
validated on the second one with the phantom
repositioned at nominally the same locations $q=1$--$4$ and $q=9$. Unlike the synthetic case, where training and validation differ by target location, the validation measurements differ here in acquisition conditions and in the actual target placement.} For
each measurement, the supervisory contrast $\mathbf{z}^{*}_i$ is
synthesized from a parametric body model with relative permittivity
contrast $\Delta\varepsilon = 1.53$ (soft tissue/muscle at $2.45$\,GHz),
yielding $\sim\!6300$ active voxels per scene (about $0.74\%$ occupancy
of the imaging grid $N$).

The measurement scenarios corresponding to the reported reconstructions are illustrated in Fig.~\ref{fig:meas_scenario}, showing respectively the single-target and two-target configurations used in the experiments.
The next sections compare W-LISTA and LoRaW-LISTA reconstruction strategies, analyze the reconstructed MIP image quality and the inference time per voxel considering different high-performance computing hardware.

\begin{table}[t]
	\centering
	\caption{MIP image quality metrics (SCR, NCC, SSIM) on the real indoor measurements: mean over the validation measurements ($q=1,2,3,4,9$) and,
in parentheses, value at $q=9$; best value per metric in bold.}
	\label{tab:metrics_real}
	\setlength{\tabcolsep}{5pt}
	\begin{tabular}{lccc}
		\toprule
		\textbf{Method} & \textcolor{black}{\textbf{SCR}} & \textbf{NCC} & \textbf{SSIM} \\
		\midrule
		\textbf{MF}              & $3.03\,(3.62)$ & $0.605\,(0.673)$ & $0.017\,(0.029)$ \\
		\textbf{ISTA}            & $3.27\,(4.00)$ & \textcolor{black}{$0.625\,(0.715)$} & $0.017\,(0.029)$ \\
		\textcolor{black}{\textbf{LISTA}} & \textcolor{black}{$3.28\,(4.12)$} & \textcolor{black}{$0.579\,(0.698)$} & \textcolor{black}{$0.020\,(0.028)$} \\
		\textbf{W-LISTA}         & \textcolor{black}{$4.76\,(7.33)$} & \textcolor{black}{$\mathbf{0.626}\,(\mathbf{0.716})$} & \textcolor{black}{$0.064\,(0.164)$} \\
		\textbf{LoRaW-LISTA}     & $\mathbf{5.56}\,(\mathbf{8.78})$ & $0.620\,(0.697)$ & $\mathbf{0.110}\,(\mathbf{0.272})$ \\
		\bottomrule
	\end{tabular}
\end{table}

\begin{table}[t]
	\centering
	\caption{Inference time per cube meter reconstruction volume (sec./m$^3$) and model footprint (byte), on 3 example GPUs.}
	\label{tab:inference_time}
	\setlength{\tabcolsep}{5pt}
	\small
	\begin{tabular}{lcccc}
		\toprule
		& \multicolumn{3}{c}{\textbf{Time} [sec./m$^3$]} & \textbf{Size} \\
		\cmidrule(lr){2-4}
		\textbf{Method} & RTX A6000 & RTX 3090 & RTX 5090 & [MB] \\
		& {\footnotesize 768\,GB/s} & {\footnotesize 936\,GB/s} & {\footnotesize 1792\,GB/s} & \\
		\midrule
		MF          & $2.3$--$2.9$ & $1.2$--$2.3$   & $0.7$--$1.2$   & $<0.01$ \\
		ISTA        & $47.5$--$48.7$ & $20.9$--$25.5$ & $16.2$--$17.4$ & $<0.01$ \\
		W-LISTA     & $48.7$--$51.0$ & $25.5$--$27.8$ & $16.2$--$18.6$ & $0.012$ \\
		LoRaW-LISTA & $48.7$--$51.0$ & $25.5$--$27.8$ & $16.2$--$18.6$ & $1.67$ \\
		\bottomrule
	\end{tabular}
\end{table}

\subsection{W-LISTA and LoRaW-LISTA analysis}
The reported Figs.~\ref{fig:real_cfr_epoch} and~\ref{fig:real_cfr} illustrate the impact of sparsity-promoting spatial regularization on the reconstructed map as in~\eqref{mip} and the interpretability of holographic RF images. Fig.~\ref{fig:real_cfr_epoch} shows the W-LISTA and LoRaW-LISTA reconstructions for a target placed at position $q=9$ (central position) across successive training epochs, allowing a visual comparison of the two methods on the same well-illuminated configuration. Fig.~\ref{fig:real_cfr} extends the comparison to $4$ distinct target positions, namely position $q=1,2,3,4$, with ground-truth location shown on the left, reporting for each the reconstructed occupancy obtained with MF back-projection, ISTA, W-LISTA and LoRaW-LISTA. In all cases the unrolled networks recover a sharper and better-localized body occupancy than the MF and ISTA baselines, with LoRaW-LISTA further refining the reconstructed map.

\textcolor{black}{Table~\ref{tab:metrics_real} reports the same metrics on the real measurements, with the addition of the Normalized Cross-Correlation (NCC), all methods being evaluated at the same training epoch. Plain LISTA performs similarly as ISTA: retuning the step-size and threshold schedule gives minor improvements. W-LISTA and LoRaW-LISTA outperform all baselines in SCR and SSIM, confirming that the learned spatial $\ell_1$ regularization concentrates the reconstructed energy on the body support and better preserves its structure. The low-rank correction of LoRaW-LISTA further improves over W-LISTA: the SCR increases by about $17\%$ on average ($20\%$ at the central position)
and the SSIM by about $70\%$, an improvement that concerns again the shape of the reconstructed support, and is obtained by compensating the residual model mismatch of the real data. In contrast, the NCC metric is far less discriminative, remaining similar for all methods: it responds mainly to the localization of the target, which every method already recovers correctly.}





\subsection{Computational complexity and edge deployment}
\label{sec:inference_time}

\textcolor{black}{This section quantifies the inference (reconstruction) time of the proposed
holographic pipelines across computing platforms spanning different
memory-bandwidth classes. The measurements reported here are obtained on
desktop-class GPUs, while the same analysis extrapolates, through the memory
bandwidth, to the embedded accelerators typically available at the network
edge.} Table~\ref{tab:inference_time} reports the reconstruction time normalized per m$^3$ of scene, obtained by dividing the full-volume time by the imaging-grid voxels (i.e., $32{,}768$ voxels/m$^3$). This metric is independent of the monitored volume and the chosen spatial resolution, and is therefore directly comparable across different hardware platforms.

The cost of all methods is dominated by the on-the-fly evaluation of the holographic operator, $\mathcal{O}(MN)$. MF requires a single adjoint multiplication, $\mathcal{O}(MN)$, whereas ISTA and the unrolled networks perform $2K$ such products, $\mathcal{O}(KMN)$ with $K=10$ layers. The spatial weights of W-LISTA and the rank-$r$ correction of LoRaW-LISTA add a negligible cost. LoRaW-LISTA improves reconstruction quality over W-LISTA (Table~\ref{tab:metrics_real}) at a slightly larger footprint, while MF is much faster but only useful for coarse localization.

The reconstruction time scales with the TPU/GPU memory bandwidth. This directly sets the hardware requirement for near-real-time operation: reconstructing $1$\,m$^3$ in about $30$\,sec. calls for a memory bandwidth of at least $\sim800$\,GB/s, well within reach of commercial GPUs (i.e., RTX~3090). \textcolor{black}{Embedded accelerators, of the kind considered for edge and IoT deployments, offer lower but steadily increasing bandwidths, i.e., DGX Spark ($273$\,GB/s) or AGX Orin ($\sim205$\,GB/s), so that the reconstruction of a $1$\,m$^3$ volume in a few minutes remains feasible, and near-real-time operation is still attainable on smaller sub-volumes. The inference cost is therefore set by a single, platform-independent figure of merit, namely the achievable memory bandwidth, rather than by the number of learned parameters.}

\section{Conclusions and future directions} 

This paper introduced a physics-informed algorithm-unrolling framework for large-scale 3D indoor RF holographic imaging. Two architectures were proposed: W-LISTA, which replaces the isotropic $\ell_1$ penalty of ISTA/LISTA with a
factorized spatial prior; and LoRaW-LISTA, which additionally learns a low-rank correction of the holographic operator to compensate for the residual mismatch of the linearized Born model. On full-wave EM simulations, W-LISTA produced cleaner, better-localized
occupancy maps than the MF and ISTA baselines, consistently improving the
error on the occupied voxels, even where the full-volume metric, dominated by
the empty background, marginally favors ISTA. On real measurements at $2.45\,$GHz, both networks sharpened the target
against the background clutter and better matched the true body shape.
\textcolor{black}{Over the ISTA and LISTA baselines, the Signal-to-Clutter Ratio (SCR) improves by about $45\%$ with W-LISTA and $70\%$ with LoRaW-LISTA. The additional gain of LoRaW-LISTA indicates that the residual bias of the linearized EM model remains the limiting factor once the structured sparsity of the prior has been learned.} The
reconstruction cost is governed by the platform memory bandwidth, so that $1$\,m$^3$ of scene is
reconstructed in less than $30$\,s on commercial GPUs.

Several directions remain open. \textcolor{black}{The learned spatial prior of W-LISTA is, by construction, tied to the region
explored during training. The low-rank correction of LoRaW-LISTA, instead, adapts the forward operator through a correction shared across all scenes, which  does not encode target locations; it is therefore a promising way to reduce the dependence on position-specific training data.} The
low-rank correction can be also read as an explicit rule that
diagnoses and repairs the mismatch between the physical model (the EM forward
model) and reality, however this is currently model-agnostic. A natural
evolution is a system that reasons about the inadequacy of the model and
selects the one to apply from a library of structured corrections, opening the
door to emerging neuro-symbolic approaches. 
Further open directions include the strong-scattering regime, where
the first-order Born approximation breaks down, the reconstruction of multiple
co-present targets, and the integration of multi-frequency measurements to
improve depth resolution.

\bibliographystyle{IEEEtran}
\bibliography{references_merged}

\vspace{-1.3cm}

\end{document}